\documentstyle[12pt]{article}
\begin{document}
\begin{titlepage}

\baselineskip 24pt plus 2pt minus 2pt

\rightline{BU-HEP-96-10 }
\vspace*{0.2cm}

\begin{center}
{\Large \bf QCDF90: a set of Fortran~90 modules for a high-level, 
efficient implementation of QCD simulations}
\end{center}

\vspace*{0.8cm}
\centerline{
{\bf Indranil Dasgupta}\footnote{e-mail: dgupta@budoe.bu.edu},
{\bf Andrea Ruben Levi}\footnote{e-mail: leviar@budoe.bu.edu},
{\bf Vittorio Lubicz}\footnote{e-mail: lubicz@weyl.bu.edu} }
\centerline{and} 
\centerline{{\bf Claudio Rebbi}\footnote{e-mail: rebbi@bu.edu} }
\centerline{\it Department of Physics, Boston University}
\centerline{\it 590 Commonwealth Avenue, Boston, MA 02215, USA}
\centerline{May 8, 1996}

\vspace*{1.5cm}
\begin{abstract}

We present a complete set of Fortran 90 modules that can be used to 
write very compact, efficient, and high level QCD programs. The modules 
define fields (gauge, fermi, generators, complex, and real fields) 
as abstract data types, together with
simpler objects such as SU(3) matrices or color vectors. 
Overloaded operators are then defined to perform all possible operations
between the fields that may be required in a QCD simulation. 
QCD programs written using these modules need not have cumbersome 
subroutines and can be very simple and transparent. 
This is illustrated with two simple example programs. 

\end{abstract}

\vfill
\end{titlepage}

\baselineskip 18pt plus 2pt minus 2pt
\setcounter{page}{1}

\section*{ \ \ PROGRAM SUMMARY}

\noindent
{\it Title of program:} QCDF90



\noindent
{\it Computer for which the program is designed:} Any computer

\noindent
{\it Computers under which the program has been tested:} 
Silicon Graphics Indigo and PowerChallengeArray, and IBM R6000 58H 
{\it Installations:} Boston University, Center for Computational
Science and Department of Physics. 

\noindent
{\it Operating systems under which the program has been tested:} IRIX
6.1, and AIX.

\noindent
{\it Programming language used:} Fortran 90

\noindent
{\it No.~of bits in a word:} 64

\noindent
{\it No.~of lines in distributed program, including test data, etc:} 7806

\noindent
{\it Keywords:} QCD, lattice gauge theory

\noindent
{\it Nature of physical problem:} Non-perturbative 
computations in QCD

\noindent
{\it Memory required to execute with typical data:} 
Varies according to the applications.
Scales proportionally to the lattice volume $NX*NY*NZ*NT$.
On a $16^4$ lattice, the example codes quenched.f90 
and propagator.f90 use
approximately 110 Mbytes and 140 Mbytes respectively.

\noindent
{\it Typical running time:} Varies according to the applications.
The example codes quenched.f90 and propagator.f90 take
approximately $45$ microsec to update an SU(3) link,
$8$ microsec to calculate a plaquette, 
and $20$ microsec for a CG step per link, using a $16^4$ lattice,
on an SGI Power-Challenge per node.

\vspace*{2.0cm}
\section*{ \ \ LONG WRITE-UP}

\section{Introduction}
\label{intro}

The computer simulation of quantum fluctuations (cf.~for instance
\cite{cjr}, \cite{rebbia}, \cite{creutza}) has been one
of the most powerful tools for obtaining information about the
non-perturbative properties of quantum field theories in general,
and, especially, of Quantum Chromodynamics (QCD) (good accounts
of progress in this field of research can be found in the proceedings of 
the yearly international symposia on lattice gauge theories \cite{latconf}; 
see also \cite{rebbib}).  These simulations, which deal with matrix and
vector fields defined over a four-dimensional space-time lattice,
involve huge number of variables and are very demanding in computer
resources.  Therefore, good payoffs can be obtained in this
domain of applications from the development of highly-efficient
code.  On the other hand, even greater
gains can be achieved through the invention of better algorithms,
which is made much easier by the availability
of high-level, structured programming tools.  
High-level programming tools are also invaluable for extracting
physical results from the data collected in the simulations,
which typically requires experimenting with different types of 
data analysis and involves substantial amounts of code development.

With the twofold goal of facilitating the development of algorithms
and applications for lattice QCD,
and of maintaining good code performance, we have taken advantage of
the possibilities offered by Fortran90 to write a set of modules for
a high-level, yet efficient implementation of QCD simulations.
Our end product is described in this long write-up, whose main
purpose is to provide researchers with all the information needed
to use our modules.  Since this effectively makes the long write-up
a reference document, it is indeed, and necessarily, ``long''.
We have nevertheless striven to be concise, in order to save space
and, especially, because we felt that a concise document would make it
easier for the user to find the relevant information.  Most of the
times the functionality provided by our modules will be obvious.
For instance, if {\tt f1} and {\tt f2} are two variables of type
fermi\_field (see later for the precise definition), {\tt f1+f2}
will have as components the sum of the components of the two fields.
Similarly, if {\tt g1} and {\tt g2} are variables of type gauge\_field,
{\tt g1*g2} will have as components the matrix products of the components
of {\tt g1} and {\tt g2}. In other
instances, however, we had to use a bit of creativity in adapting
the symbols of the language to the definitions of some further useful 
overloaded operators.  Thus, if {\tt f1} and {\tt f2} are again variables 
of type fermi\_field, {\tt f1//f2} will be for us a variable of 
type gauge\_field having for components the dyadic formed by the 
vector components of the two fermi\_fields.  For all these less
obvious definitions, there is no substitute to reading the
sections of this article, where all of our overloaded operators
are carefully documented.  

We expect that most of the users of our modules will be practitioners of
lattice gauge theory, and as such already quite knowledgeable about 
the type of variables that enter QCD simulations.  
Having in mind, however, that some of the
users might be application scientists called on to benchmark code
with which they are not too familiar, we decided to include in this
write-up a very concise description of the data structures encountered in
QCD simulations.  This is presented in the next section, which summarizes
the kinematics that has been used for lattice QCD since the pioneering 
work of Wilson \cite{wilson}.  The section that follows discusses
the all-important notion of parallel transport in presence of a gauge
field and our implementation of parallel transport via a generalization
of the C-shift operation, which we call ``U-shift''.  Sections \ref{evenand},
\ref{types} and \ref{programmingand} deal with algorithmic issues related 
to the ordering of the data, with the description of the data types,
and with some further considerations of programming and efficiency.
The remaining sections are devoted to a detailed description of
our modules and of the functionality which they provide. 

\section{Geometry and variables}
\label{geometryand}

We consider a four-dimensional lattice with extent $ \tt NX, NY, NZ$ and 
$\tt NT$ in the four directions.  
We will assume that $\tt NX,NY,NZ,NT$ are all even.
A lattice site will be labeled by four integer valued variables
$\tt x,y,z,t$ with
\begin{equation}
\tt 0 \le x < NX, \quad 0 \le y < NY, \quad 0 \le z < NZ, 
\quad 0 \le t < NT \ .
\label{sites}
\end{equation}
When convenient, we will denote the collection of these four indices
by ${\bf x}$.

We will assume periodic boundary conditions.

The physical system is defined in terms of two types of variables
(also called the dynamical variables):
the gauge fields and the Fermi fields.

The components of a gauge field are $3 \times 3$ unitary, unimodular matrices
(i.e.~elements of the group $SU(3)$, the so called ``color'' group) defined
over the oriented links of the lattice.  Later we will see that programming
considerations demand a more involved layout of data, but, conceptually,
a gauge field can be considered as a multidimensional array of complex
variables
\begin{equation}
\tt U(3,3,0:NX-1,0:NY-1,0:NZ-1,0:NT-1,4) \ ,
\label{gaugefield}
\end{equation}
where the first two indices of the generic array element
$\tt U(i,j,x,y,z,t,m)$ are the indices of the $SU(3)$ matrix, whereas
$\tt x,y,z,t$ label a lattice site and $\tt m=1 \dots 4$ labels one of the
four lattice links having origin at the site and oriented in the
positive $\tt m$ direction.  When convenient we will use the more
compact notation $U_{ij,{\bf x}}^{\mu}$ to denote the gauge
field elements, or $U_{\bf x}^{\mu}$ to denote the whole matrix defined
over the link (in this compact notation we follow the common practice of
using a Greek letter to denote the direction
of the link).  Another useful notation consists in representing by
$\hat\mu$ a four-vector having its $\mu$ component equal to 1,
all other components equal to zero.  With this notation, one can say
that the gauge variable $U_{\bf x}^{\mu}$ is defined over the
oriented link from ${\bf x}$ to ${\bf x}+{\hat\mu}$.

The components of a Fermi field are defined over the sites of the
lattice.  They are 3-dimensional complex vectors with respect to the
matrices of the color group and carry an additional spin index 
$\tt s$ ranging from 1 to 4.  Thus the data layout of a Fermi
field can be represented conceptually in terms of an array
of complex variables
\begin{equation}
\tt f(3,0:NX-1,0:NY-1,0:NZ-1,0:NT-1,4) \ ,
\label{fermifield}
\end{equation}
where the first index of the generic array element $\tt f(i,x,y,z,t,s)$
is the color index, $\tt x,y,z,t$ label the site and $\tt s$ is the spin index.
When convenient we will use a more compact notation $\psi_{i,{\bf x},s}$
for the components of a Fermi field, or $\psi_{{\bf x},s}$ for
the whole color vector, or even just $\psi_{\bf x}$.
 
In the field theoretical definition of the physical system the components
of a Fermi field would be anticommuting elements of a Grassmann algebra
with integration.  The rules of integration over elements of a Grassmann
algebra $\psi_a$, $\bar\psi_b$ ($a$ and $b$ stand for any complete
set of indices) have, as their most important consequence, the formula
\begin{equation}
\int \prod_a (d\bar\psi_a \,d\psi_a) \exp[\sum_{a,b} \bar\psi_a 
A_{a,b} \psi_b] = {\rm Det}[A] \ .
\label{gaussianint}
\end{equation}
In computational applications ${\rm Det} [A]$ or its derivatives with
respect to the dynamical variables are calculated by means of ordinary
complex variables $\phi_a$, $\bar\phi_b$ making use of the identity
\begin{equation}
{\rm Det}[A] = \int \prod_a ({d\bar\phi_a \,d\phi_a \over \pi})
\exp[\sum_{a,b} \bar\phi_a [A^{-1}]_{a,b} \phi_b]  \ .
\label{gaussiancplx}
\end{equation}
Thus effectively one deals with arrays of complex variables as in
(\ref{fermifield}).

\section{The notion of U-shift}
\label{thenotion}

The gauge field serves to define the transport of dynamical variables
between neighboring sites.  Gauge theories are characterized by the
property of local gauge invariance.  In the present context this means
that it is always possible to redefine the Fermi variables by an $SU(3)$
transformation
\begin{equation}
\psi_{i,{\bf x},s} \to \psi'_{i,{\bf x},s} =
\sum_j G_{ij,{\bf x}} \psi_{j,{\bf x},s}  \ ,
\label{gaugetrfm}
\end{equation}
where the elements of the gauge transformation $G_{ij,{\bf x}}$ are
$SU(3)$ matrices defined over the sites.  All of the physical quantities
must remain invariant under such transformations.

It is clear that, if the Fermi fields transform according to (\ref{gaugetrfm}) 
with a $G_{ij,{\bf x}}$ that changes from site to site, a straightforward
finite difference (as one would use in the approximation of a derivative)
\begin{equation}
(\Delta \psi)_{i,{\bf x},s} = \psi_{i,{{\bf x}+\hat\mu},s} -
\psi_{i,{\bf x},s}  \ 
\label{finitediff}
\end{equation}
will produce meaningless results.  Rather, one should ``transport''
the variable $\psi_{{\bf x}+\hat\mu}$ from the site 
${{\bf x}+\hat\mu}$ to the site $\bf x$ by means of the gauge variable
$U_{\bf x}^{\mu}$ defining a shifted variable 
\begin{equation}
\psi^{\rm shifted}_{i,{\bf x},s} = \sum_j U_{ij,{\bf x}}^{\mu}
\psi_{j,{{\bf x}+\hat\mu},s} 
\label{shiftfm}
\end{equation}
and then define a gauge covariant finite difference
\begin{equation}
(D \psi)_{i,{\bf x},s} = \psi^{\rm shifted}_{i,{\bf x},s} - 
\psi_{i,{\bf x},s}  \ .
\label{gfinitediff}
\end{equation}
Under a gauge transformation the gauge field itself changes according
to 
\begin{equation}
U_{ij,{\bf x}}^{\mu} \to U_{ij,{\bf x}}^{\prime \mu} =
\sum_{kl} G_{ik,{\bf x}} U_{kl,{\bf x}}^{\mu} [G^{-1}]_{lj,{\bf x}+\hat\mu}
 \ .
\label{gaugetru}
\end{equation}
 From Eqs.~(\ref{gfinitediff},\ref{gaugetrfm},\ref{gaugetru}) one can verify
that under a gauge transformation the gauge covariant finite difference 
changes like $\psi$ itself:
\begin{equation}
(D \psi)_{i,{\bf x},s} \to (D \psi')_{i,{\bf x},s} =
\sum_j G_{ij,{\bf x}} (D \psi)_{j,{\bf x},s}  \ .
\label{gaugetrD}
\end{equation}
Thus the gauge covariant finite difference is a meaningful construct
and quantities such as its magnitude or the scalar product
$\sum_{i} \bar\psi_{i,{\bf x},s} (D \psi)_{i,{\bf x},s'}$ are gauge invariant
and thus physically well defined. 

It is clear from the above that a circular shift (C-shift) of an array
such as $\tt f(3,0:NX-1,0:NY-1,0:NZ-1,0:NT-1,4)$ will generally be
complemented by multiplication by an element of the gauge field.
We will therefore define a U-shift operation in the following manner.

A U-shift with positive direction parameter $\mu = 1 \dots 4$ of the
Fermi field $\psi_{i,{\bf x},s}$ produces the array 
$\psi^{\rm shifted}_{i,{\bf x},s}$ as given by Eq.~(\ref{shiftfm}).

A U-shift with negative direction parameter 
$\mu' = -\mu = -1 \dots -4$ of the
Fermi field $\psi_{i,{\bf x},s}$ produces the array 
\begin{equation}
\psi^{\rm shifted}_{i,{\bf x},s} = \sum_j U_{ij,{\bf x}-\hat\mu}^{\dagger \mu}
\psi_{i,{{\bf x}-\hat\mu},s} \ ,
\label{negshiftfm}
\end{equation}
this latter equation being motivated by the fact that the transport
factor over a link crossed in the negative direction is the 
Hermitian adjoint (or equivalently the
inverse, with a unitary group $U^{\dagger}=U^{-1}$)
of the transport factor for the positively oriented link.

We define a U-shift for the gauge field as well.  Since the gauge field
elements have two color indices which should be associated with the
beginning and end of the link (cf.~the gauge transformation properties
of the gauge field variables Eq.~(\ref{gaugetru})) the U-shift of a gauge
field will involve two matrix multiplications.  Moreover, it will be
convenient to define its action on a generic gauge field, denoted below by $V$,
not necessarily identical to $U$.  The idea is that in general there  
will be several variables with the properties of a gauge field (see 
the type definitions below) but there will always be one
well defined ``master gauge field'', denoted by $U$, which will serve
to define the transport of all gauge dependent variables.  With this
in mind the action of a U-shift on a gauge field is defined as follows.

A U-shift with positive direction parameter $\mu = 1 \dots 4$ of the
gauge field $V_{ij,{\bf x}}^{\nu}$ produces the array 
\begin{equation}
V^{{\rm shifted},\nu}_{i,{\bf x}} = \sum_{kl} U_{ik,{\bf x}}^{\mu}
V_{kl,{{\bf x}+\hat\mu}}^{\nu} U_{lj,{\bf x}+\hat\nu}^{\dagger \mu}\ .
\label{shiftv}
\end{equation}

A U-shift with negative direction parameter 
$\mu' = -\mu = -1 \dots -4$ of the
gauge field $V_{ij,{\bf x}}^{\nu}$ produces the array 
\begin{equation}
V^{{\rm shifted},\nu}_{i,{\bf x}} = 
\sum_{kl} U_{ik,{\bf x}-\hat\mu}^{\dagger \mu}
V_{kl,{{\bf x}-\hat\mu}}^{\nu} U_{lj,{\bf x}-\hat\mu+\hat\nu}^{\mu}\ .
\label{negshiftv}
\end{equation}

When acting on field variables which carry no color index (we will define
such field variables below) the U-shift reduces to an ordinary C-shift.

\section{Even and odd components of field variables}
\label{evenand}

All lattice sites can be subdivided into ``even'' and ``odd'' sites
according to whether the sum of the integer valued coordinates
$\tt x+y+z+t$ is even or odd (checkerboard subdivision).  
Correspondingly all field variables can
be divided into even and odd variables (for a gauge field variable
we base the subdivision on the origin of the link over which the variable 
is defined, i.e. the $\tt x,y,z,t$ indices of the array~(\ref{gaugefield})).
With periodic boundary conditions and with an even lattice size in
all directions, a C-shift or a U-shift of an even field variable produces
an odd field variable and vice versa.  There are many algorithms which
demand, especially in the context of a parallel implementation, that
even and odd variables be treated separately.  For example, in a Monte
Carlo simulation algorithm all variables at even sites can be upgraded
simultaneously while those at odd sites are kept fixed and vice versa.
We will accommodate these algorithmic demands by defining all of
our field variables as arrays of even or odd field variables.
We will do so by taking advantage of the type definition as follows.
All field variables will be defined through a type.  The first component
of the type will be an integer variable $\tt parity$ which will take
values $\tt 0$ and $\tt 1$ for variables defined over even and odd 
sites respectively.
It will also be convenient to use the value $\tt -1$ to characterize
a field with parity undefined.

For the gauge variables it will be convenient
to include in the type a single $\mu$ component (of definite parity,
of course).  Thus, in addition to the variable $\tt parity$,
the type will contain an integer variable $\tt dir$, 
taking values $\tt 1 \dots 4$, to denote the direction of the link 
(i.e.~the value of the index $\mu$).  It will be convenient to let
$\tt dir$ also take the value $\tt 0$, to characterize $3 \times 3$
complex matrices which are defined over the sites rather than
over the links, such as the $SU(3)$ matrices of the gauge transformation
in Eq.~(\ref{gaugetrfm}).

Finally the type will then contain an array, denoted by $\tt fc$ 
(for field component) which will contain all the field variables
defined over the sites of a given parity.

Insofar as the indexing of the array is concerned, this is to a large
extent arbitrary, provided that the mapping between the array indices
and the actual Cartesian coordinates of the site is well defined.
For instance, one could collapse two neighboring ``time'' slices
into a single one and use indices $\tt x,y,z,t$ where $\tt t$ ranges
now from $\tt 0$ to $\tt NT2-1=NT/2-1$.  On the other hand,
with many architectures efficiency considerations recommend that 
the indices $\tt x,y,z,t$ be fused into a single index,
spanning the range $\tt 0$ to $\tt NXYZT2-1=NX*NY*NZ*NT/2-1$.
This will typically be the case when, because of a vectorized or
superscalar architecture, the instructions are pipelined and
longer arrays give rise to better performance.  In principle
a good optimizing compiler should recognize when the individual 
loops over the $\tt x,y,z$ and $\tt t$ indices can be fused
into a single one and take advantage of this possibility.  However,
some compilers may be able to fuse only a limited number of nested loops
or, alternatively, this type of optimization may be hindered by the
the presence of further indices or by a large number of instructions
within the loops.  Since the use of types and overloaded
operators makes the actual indexing of the arrays transparent
to the user, we decided to use a single index to label all of
the sites of a definite parity.  This index is constructed by
going through the sites of definite parity in a lexicographic
order, increasing the $x$ coordinate first, then $y$, $z$ and $t$,
but, as stated above, the ordering of the sites is largely immaterial.
For all those operations which are performed locally over the sites,
the detailed mapping between the index and the geometry of the lattice
is clearly irrelevant.  It is, of course, of consequence for
the implementation of the shift operations and for accessing
the component of a field at a definite Cartesian site.
For such purposes we provide the specifics of the mapping through
some global variables and an initialization subroutine.
We define the following global variables:

{\leftline{\tt INTEGER, DIMENSION(0:NX-1,0:NY-1,0:NZ-1,0:NT-1) :: 
xyzt\_index}}
{\leftline{\tt INTEGER, DIMENSION(0:NX-1,0:NY-1,0:NZ-1,0:NT-1) :: 
xyzt\_parity}}
{\leftline{\tt INTEGER, DIMENSION(0:NXYZT2-1,0:1,4) :: 
xyzt\_cartesian}}
{\leftline{\tt INTEGER, DIMENSION(0:NXYZT2-1,0:1,8) ::
xyzt\_neighbor}}
{\leftline{\tt LOGICAL shift\_initialized}}

{\parindent=0pt where the parameter $\tt NXYZT2=NX*NY*NZ*NT/2$ 
equals one half of the total number of sites in the lattice.}

The arrays defined above are initialized by executing the subroutine

{\parindent=0pt {\tt shift\_initialization}.  
The variable {\tt shift\_initialized}
is initialized to {\tt .FALSE.}  All of the function calls
which implement the overloaded shift operators check the value
of  {\tt shift\_initialized}.  If this is {\tt .FALSE.}, the
subroutine {\tt shift\_initialization} is called and the arrays are 
properly initialized.  Before returning, {\tt shift\_initialization} sets
{\tt shift\_initialized} to {\tt .TRUE.}  From this moment on
the arrays can be used to establish the mapping between the
Cartesian coordinates and the indices within the sublattices
of definite parity.  The programmer wishing to use these arrays
before any shift operation is performed can, of course, initialize
them directly via a call to {\tt shift\_initialization}. }

The array component {\tt xyzt\_index(x,y,z,t)} gives the index of the
field component defined over the site with Cartesian coordinates
$\tt x,y,z,t$. 

{\parindent=0pt  {\tt xyzt\_parity(x,y,z,t)} gives the parity of the
site ($\tt xyzt\_parity(x,y,z,t)$
$\tt  = x+y+z+t\;  MOD \; 2$).
{\tt xyzt\_cartesian(i,p,m)} gives the Cartesian coordinate ({\tt x,y,z,t}
for {\tt m=1,2,3,4} respectively) of the site with index {\tt i} and 
parity {\tt p}. Finally {\tt xyzt\_neighbor(i,p,m)} gives the index
of the nearest neighbor site in direction {\tt m} of a site with index
{\tt i} and parity {\tt p}.  The convention is that the
values {\tt m=1,2,3,4}
correspond to the nearest neighbor in the forward $x,y,z,t$ 
directions, whereas {\tt m=5,6,7,8} correspond to the 
nearest neighbor in the backward $x,y,z,t$ directions, respectively.}
 
\section{Types}
\label{types}

We define the following F90 types.

\subsection{Type gauge\_field}
\label{typeg}

\vskip 4mm
{\baselineskip 5mm \tt
\leftline{TYPE gauge\_field}
\leftline{\quad INTEGER parity}
\leftline{\quad INTEGER dir}
\leftline{\quad COMPLEX(REAL8),DIMENSION(3,3,0:NXYZT2-1)::fc}
\leftline{END TYPE}
}

As discussed in the previous section, a variable of type gauge\_field 
contains the components of a gauge field defined over all the links
of direction $\tt dir$ emerging from the lattice sites  
of a given $\tt parity$.  The field component $\tt fc(i,j,xyzt)$
is an array of double precision complex variables (the kind $\tt REAL8$ is
defined in the module ``precisions'', see below), where 
$\tt i,j$ are the indices of the $SU(3)$ matrix and $\tt xyzt$ labels
the site within the subset of sites of a definite parity.

\subsection{Type full\_gauge\_field}  
\label{typeu}

\vskip 4mm
{\baselineskip 5mm \tt
\leftline{TYPE full\_gauge\_field}
\leftline{\quad TYPE(gauge\_field), DIMENSION(0:1,4) :: uc}
\leftline{END TYPE}
}

A variable of type full\_gauge\_field is meant to store an entire
gauge field configuration, i.e.~8 variables of type gauge\_field
corresponding to the two parity components and the 4 direction
components of a full gauge field.  Although the $\tt parity$
and $\tt dir$ components of the individual $\tt uc(i,j)$ components
can be given any value, good programming practice recommends that one sets
$\tt uc(i,j)\%parity = i $, $\tt uc(i,j)\%dir = j $.

\subsection{Type fermi\_field}        
\label{typef}
                                         
\vskip 4mm
{\baselineskip 5mm \tt
\leftline{TYPE fermi\_field}
\leftline{\quad INTEGER parity}
\leftline{\quad COMPLEX(REAL8),DIMENSION(3,0:NXYZT2-1,4)::fc}
\leftline{END TYPE}
}

A variable of type fermi\_field 
contains the components of a Fermi field defined over the lattice sites  
of a given $\tt parity$.  The field component $\tt fc(i,xyzt,s)$
is an array of double precision complex variables, where 
$\tt i$ is the color index, $\tt xyzt$ labels
the site and $\tt s$ is the spin index of the field.

\subsection{Type complex\_field}      
\label{typec}
                                         
\vskip 4mm
{\baselineskip 5mm \tt
\leftline{TYPE complex\_field}
\leftline{\quad INTEGER parity}
\leftline{\quad COMPLEX(REAL8),DIMENSION(0:NXYZT2-1)::fc}
\leftline{END TYPE}
}

The type complex\_field is introduced to store an array of complex
numbers $\tt fc(xyzt)$ defined over the lattice sites
of a given $\tt parity$. Although one could also store such variables
in an array of complex numbers, defining a type has the advantage
that one can record the parity of the field and that it becomes possible
to define overloaded operators (intrinsic operations on intrinsic
types cannot be overloaded).  A similar remark applies to the
type real\_field defined below.

\subsection{Type real\_field}         
\label{typer}
                                         
\vskip 4mm
{\baselineskip 5mm \tt
\leftline{TYPE real\_field}
\leftline{\quad INTEGER parity}
\leftline{\quad REAL(REAL8),DIMENSION(0:NXYZT2-1)::fc}
\leftline{END TYPE}
}

The type real\_field is introduced to store an array of real
numbers $\tt fc(xyzt)$ defined over the lattice sites
of a given $\tt parity$.

\subsection{Type generator\_field}    
\label{typege}
                                         
\vskip 4mm
{\baselineskip 5mm \tt
\leftline{TYPE generator\_field}
\leftline{\quad INTEGER parity}
\leftline{\quad INTEGER dir}
\leftline{\quad REAL(REAL8),DIMENSION(8,0:NXYZT2-1)::fc}
\leftline{END TYPE}
}

Although for computational purposes it is useful to store the components
of an $SU(3)$ gauge field as $3 \times 3$ complex matrices, a general
$SU(3)$ matrix is a function of only 8 real independent parameters. 
In particular, given an 8-dimensional real vector with components
$v_k$ one can associate to it the $SU(3)$ matrix
\begin{equation}
U_{ij}= \bigg[\exp \big( \imath \sum_{k=1}^8  v_k \lambda^k \big)
 \bigg]_{ij} \ , 
\label{expv}
\end{equation}
where the matrices $\lambda^k$ form a basis in the space 
of Hermitian traceless $3 \times 3$ matrices and satisfy the
equations ${\rm Tr} (\lambda^k  \lambda^{k'}) = 0$ for
$k \ne k'$, ${\rm Tr} (\lambda^k)^2 = 2$.

The term group generator is commonly used to refer to a traceless
Hermitian matrix, such as 
\begin{equation}
H_{ij}= \sum_{k=1}^8  v_k \lambda^k_{ij} 
\label{generator}
\end{equation}
in Eq.~(\ref{expv}).  For some algorithms it is convenient to deal
directly with the components $v_k$ of a generator, rather with
the exponentiated matrix $U$ or the Hermitian matrix $H$.  For this
reason we provide the type generator\_field, aimed at storing generator
components defined over the sites of a given {$\tt parity$}.  Since generators
are frequently associated to gauge field variables, we give the
type generator\_field a $\tt dir$ component as well.

\subsection{Type matrix}              
\label{typem}
                                         
\vskip 4mm
{\baselineskip 5mm \tt
\leftline{TYPE matrix}
\leftline{\quad COMPLEX(REAL8),DIMENSION(3,3)::mc}
\leftline{END TYPE}
}

The type matrix is defined for programming convenience,
in order to allow for the overloading of operators and assignments.
For instance, it makes it possible to define an operation $\tt g*m $, 
where the variables $\tt g$ and $\tt m$
are of type gauge\_field and matrix respectively, 
which implements the matrix product of the components of a gauge
field times a constant matrix.

\section{Programming and efficiency considerations}
\label{programmingand}

\subsection{One layer versus two layer data structure}
\label{onelayer}

Conceptually our variables would be most naturally defined in terms of
a two layer data structure.  At the bottom layer we would find objects
such as a single $SU(3)$ matrix or a single color vector, i.e.~three
dimensional complex vector.  Overloaded operations such as matrix 
multiplication or multiplication of a matrix times a color vector
would also be defined. At the top layer we would then use these
objects to define extended fields, such as the gauge field, consisting
of an array of objects of type matrix.  Operators among the objects 
of the top layer would be built from the elemental operators
already defined at the bottom layer. However appealing, this organization 
of the data would almost certainly imply a huge penalty in efficiency.
It is indeed reasonable to expect that the compiler will implement
overloaded operations in terms of function calls.  In a two layer
structure, then, an operation such as the addition of two Fermi
fields would be implemented via repeated calls, site by site, to 
the function which adds the color vector components of the two
fields.  It is clear that this use of function calls
at very low granularity would imply a heavy computational burden.
The only way to regain efficiency would be to inline the function
calls implementing the elemental operations.  While in principle
this is possible, it is not reasonable to expect that compilers
would generally allow inlining of function calls that implement operations
among derived data types over which they have little direct control.
For this reason we decided to forfeit the possibility of defining
a two layer data structure, however conceptually pleasing this
may be, and 
organized all of our data into a single layer of user defined types.
Thus the types which we introduce to define
extended fields are, essentially, F90 arrays complemented with one
or two variables ($\tt parity, dir$) specifying their attributes. 
As a consequence the computational cost for the use of overloaded 
operators between our data structures should not be any bigger than
the cost of a call to a function or subroutine that manipulates large
arrays.  On the other hand, the advantages we gain in code structure
and ease of programming are truly remarkable.

\subsection{Overloaded assignments}
\label{overloaded}

The use of overloaded operators may imply the creation of more 
temporaries and, consequently, more motion of data than a straightforward
implementation of operations among arrays.  Consider for example the
following operation among variables of type fermi\_field:
\begin{equation}
\tt f1=f1+f2+f3 \ .
\label{addfff}
\end{equation}
(We will formally define the addition of Fermi fields later, but it
will perform the  obvious operation of adding the $\tt fc$ components
of the fields.)

With ordinary arrays the compiler might put the result of $\tt f1+f2$
in a temporary $\tt t1$ and then add $\tt t1$ and $\tt f3$ placing
the result in $\tt f1$.  Thus there would be two write-to-memory 
operations per
component of the arrays.  (A good optimizing compiler could even use
registers, dispensing with the creation of the temporary and 
of one of the copies to memory.)  However, if the overloaded addition 
of Fermi fields is implemented via function calls, what we expect to
happen is that the function implementing $\tt f1+f2$ places the
result into a temporary $\tt t1$ returning the address of the 
corresponding data structure to the calling program.  The compiler
at this point will probably copy $\tt t1$ into a temporary $\tt t2$, since
it would not be safe to pass the addresses of $\tt t1$ and $\tt f3$
to the add function which will likely put
the result into $\tt t1$ again.  Finally, the result will be copied
into $\tt f1$.  If implemented in this manner, the entire operation
involves four write-to-memory operations: to $\tt t1$, to $\tt t2$,
to $\tt t1$ again and to $\tt f1$. (Of course, all of the above is 
implementation dependent.  As far as we know, F90 does not specify how the 
variables should be passed in function calls.  An operating system could 
let the calling program pass to a function the address where it expects
the result, making the
call $\tt a=function(b,c)$ effectively identical to 
$\tt CALL \; subroutine(a,b,c)$. In this case 
the composite operation~(\ref{addfff})
could be implemented with two copies to memory only.)

The procedure could be drastically simplified through the use of
an overloaded assignment $\tt +=$.  Instruction~(\ref{addfff}) could be
written
\begin{equation}
\tt f1\,+=f2+f3 \ ,
\label{pefff}
\end{equation}
which the compiler would implement by issuing first a call to a function
that adds $\tt f2$ and $\tt f3$ returning the result in $\tt t1$.  The
addresses of $\tt f1$ and $\tt t1$ would then be passed to a subroutine,
e.g.~$\tt plus\_eq(a,b)$ that implements the operation $\tt f1=f1+t1$
among the components of the data types.  The required number of copies
to memory would be only two.

In order to allow for these possible gains in efficiency, we have defined
a large set of overloaded assignments, which will be detailed in the
description of the module ``assign'' given below.  Since F90 permits
only the use of the $\tt =$ symbol for the assignment, we have implemented
its overloading by defining two global variables: a character variable
$\tt assign\_type$ and an integer variable $\tt assign\_spec$ (for assign
specification, introduced to accommodate assignments of a more
elaborate nature).  The default values of these variables are $\tt '='$
and $\tt 0$.  They are initialized with these values and reset to
their default values at the end of all overloaded assignments.  We
follow this procedure to avoid the occurrence of accidental erroneous
assignments.  When $\tt assign\_type$ equals  $\tt '='$ the result of
the assignment between variables of identical type is the expected
copy of the data structure at the r.h.s. into the variable at the l.h.s..
(We also define overloaded $\tt '='$ assignments between variables
of different type; the results of such assignments are explained
in the description of the module ``assign''.)  Overloaded
assignments such as $\tt a\,+=b$ are obtained by setting 
$\tt assign\_type$ (and possibly $\tt assign\_spec$) to the appropriate
value immediately before the assignment.  We recommend the following
pattern for the instructions:
\begin{equation}
\tt assign\_type='+'; \quad a=b  
\label{asgnone}
\end{equation}
or (this implements a U-shift from direction $\tt n$)
\begin{equation}
\tt assign\_type='u'; \quad assign\_spec=n;  \quad a=b  
\label{asgntwo}
\end{equation}
The overloaded assignments are implemented via case constructs, which
make reference to the values of the global variables $\tt assign\_type$, 
$\tt assign\_spec$.  A simplified version of the code for an assignment
would be as follows:

\vskip 4mm
{\baselineskip 5mm \tt
\leftline{SUBROUTINE typea\_eq\_typeb(a,b)}
\leftline{\quad TYPE(typea), INTENT(INOUT) :: a}
\leftline{\quad TYPE(typeb), INTENT(IN) :: b}
\leftline{\quad SELECT CASE(assign\_type)}
\leftline{\quad CASE('=')}
\leftline{\quad \quad implements a=b}
\leftline{\quad CASE('+')}
\leftline{\quad \quad implements a=a+b}
\leftline{\quad CASE DEFAULT}
\leftline{\quad \quad returns an error message and stops execution 
if the value}
\leftline{\quad \quad of assign\_type does not correspond
to any defined assignment}
\leftline{\quad END SELECT}
\leftline{\quad assign\_type='='; \quad assign\_spec=0}
\leftline{END SUBROUTINE typea\_eq\_typeb}
}

We wish to emphasize that the structure of
data and operations which we have introduced may still cause loss
of efficiency with some compilers, even with an optimizing one.
It might happen that code performing the same calculations as a
code written in terms of our data structures, but formulated without
use of any derived data types, is converted, upon compilation, into a more
efficient executable.  However, we designed our data structure
and defined our operators and assignments in a way which should present
no barrier to a highly efficient, parallelizing compilation.  It will
be an interesting experiment to verify how different compilers respond
to it.

\section{Modules}
\label{modules}

\subsection{Module precisions}
\label{precisions}

This module defines two kind parameters, $\tt REAL8 $ and $\tt LONG $.
These parameters store the kind of an 8-byte floating point variable
and of an 8-byte integer variable.  They are used to render the 
kind definitions machine independent.  $\tt INTEGER(LONG) $ variables
are used only for the parallel generation of pseudorandom numbers 
in a system independent way (cf.~the module ``random\_numbers'').  
If 8-byte integers are not supported by the architecture, the module 
random\_numbers should be modified to run with shorter integers
or to use system supplied parallel pseudorandom numbers, and the 
definition of $\tt LONG$ should be changed accordingly.

\subsection{Module global\_module}
\label{global}

This module defines the integer constants $\tt NX$, $\tt NY$, $\tt NZ$
and $\tt NT$ which specify the size of the lattice. $\tt NX$, $\tt NY$, 
$\tt NZ$, $\tt NT$ must all be even.  It defines the reduced
temporal extent $\tt NT2= NT/2$, and the products 
$\tt NXYZT = NX*NY*NZ*NT$, $\tt NXYZT2 = NX*NY*NZ*NT2$. 
It also defines for convenience the constants
$\tt NCGV = 9*NXYZT2$, $\tt NCFV = 12*NXYZT2$, 
$\tt NRGV= 2*NCGV$, $\tt NRFV= 2*NCFV$, $\tt NRGEV= 8*NXYZT2$,
which are equal to the number of complex or, respectively, real variables
in the $\tt fc$ components of the types gauge\_field, fermi\_field 
and generator\_field. 

All of the types introduced in Sect.~\ref{types} are declared in this module.

Finally the module declares a few global variables, namely, the master
gauge field:

\leftline{$\tt TYPE(full\_gauge\_field)  \quad u $ }

{\parindent=0pt
the assignment variables (cf.~Sect.~\ref{overloaded}):}

\leftline{$\tt CHARACTER \quad assign\_type $} 
\leftline{$\tt INTEGER \quad assign\_spec $} 

{\parindent=0pt
the arrays {\tt xyzt\_index}, {\tt xyzt\_parity}, {\tt xyzt\_cartesian},
{\tt xyzt\_neighbor} and the logical variable {\tt shift\_initialized},
already mentioned in Sect.~\ref{evenand},}

{\parindent=0pt
the context logical array, used in conditional operations (cf.~the module
``conditionals''):}

\leftline{$\tt LOGICAL, DIMENSION(0:NXYZT2-1) :: 
\; context $}

{\parindent=0pt
and the variables used for the generation of pseudorandom numbers
(see the module ``random\_numbers''):}

\leftline{$\tt INTEGER \quad seed\_a, seed\_b $} 
\leftline{$\tt INTEGER,  DIMENSION(0:NXYZT2-1) :: 
\; seeds $} 
  
{\parindent=0pt The module contains the subroutine {\tt shift\_initialization}
(see Sect.~\ref{evenand}).}

\subsection{Module constants}
\label{constants}

This module defines some useful parameters, making them available
to all program units which use it.  Namely, the following real constants
are defined: $\tt PI$ ($\pi$),   
$\tt PI2$ ($\pi/2$), $\tt TWOPI$ ($2 \pi$),   
$\tt SQRT2$ ($\sqrt{2}$), $\tt SQRT22$ ($\sqrt{2}/2$),   
$\tt SQRT3$ ($\sqrt{3}$), $\tt SQRT33$ ($\sqrt{3}/3$),   
$\tt TWOSQRT33$ ($2 \sqrt{3}/3$), the complex constant $\tt IU$
($\imath$), and the arrays:     

\leftline{$\tt COMPLEX(REAL8), DIMENSION(3,3) :: ZERO\_m, UNIT, IU\_m$}

\leftline{$\tt COMPLEX(REAL8), DIMENSION(3) :: ZERO\_v$}

\leftline{$\tt REAL(REAL8), DIMENSION(8) :: ZERO\_ge$}

\leftline{$\tt COMPLEX(REAL8), DIMENSION(3,3,8) :: LAMBDA$}

\leftline{$\tt COMPLEX(REAL8), DIMENSION(4,4,5) :: GAMMA$}

$\tt UNIT$ and $\tt IU\_m$ are set equal to the unit matrix, 
and to $\imath$ times the unit matrix, respectively.
$\tt ZERO\_m$, $\tt ZERO\_v$, $\tt ZERO\_ge$ have all
components equal to zero.  The array $\tt LAMBDA$ stores
the components of the $\lambda$ matrices:

\leftline{ ${\tt LAMBDA(i,j,k)}=\lambda_{i,j}^k$,} 

{\parindent=0pt
and the array $\tt GAMMA$ stores the components of 
Dirac's $\gamma$ matrices,}

\leftline{${\tt GAMMA(s1,s2,m)}=
\gamma_{s{\scriptscriptstyle 1},s{\scriptscriptstyle 2}}^m,\;
m=1 \dots 5 \;,$
in our chosen representation.}

{\parindent=0pt (We follow the convention $\gamma^5 = \gamma^1 \gamma^2
\gamma^3 \gamma^4$.)}

Notice that we do not make any distinction between upper and lower indices
for the $\lambda$ and $\gamma$ matrices:  $\lambda^k=\lambda_k$,
$\gamma^m=\gamma_m$ and the use of upper or lower indices is only dictated
by notational convenience.

\subsection{Module field\_algebra}      
\label{fieldalgebra}

This module defines several overloaded operators that perform
arithmetic operations between fields and other variables.
We describe here all the operations which are defined.  For conciseness
we introduce notational conventions. We use the symbols $\tt g,$ 
$\tt u,$ $\tt f,$ $\tt c,$ $\tt r,$ $\tt ge$ and $\tt m$ 
to denote variables of type gauge\_field, full\_gauge\_field, 
fermi\_field, complex\_field, real\_field, 
generator\_field and matrix, respectively, and the
symbols $\tt complex$ and $\tt real$ to denote a complex or real variable
of kind $\tt REAL8$ (cf.~Sect.~\ref{precisions}).  When necessary, 
we will use subscripts, e.g.~$\tt f_1, f_2$, to distinguish between two 
variables of the same type.

All operators obey the following general rules:

i) When the result of the operation is a field, if the two operands
have a $\tt parity$ component, the $\tt parity$ of the result is
the $\tt parity$ of the operands if they have the same $\tt parity$,
otherwise it is undefined (i.e.~$= -1$).  If a single operand
has a $\tt parity$ component, then the $\tt parity$ of the result
takes the same value.  A similar rule applies to the direction
component of the variables of type gauge\_field and generator\_field: if 
both operands have the same $\tt dir$ or a single operand carries a $\tt dir$ 
component, then the $\tt dir$ component of the result is set to this
value.  Otherwise it is set to $0$.
 
ii) When the operator acts between fields, the operation is performed
site by site and the result is again a variable of field type.  When the
operator acts between a variable of type field and a global variable
(i.e.~$\tt m$, $\tt complex$ and $\tt real$) the site variable
is combined with the global variable.  For example, the operations
$\tt c=c_1+c_2$ and $\tt c=c_1+complex$ would be implemented as

\vskip 4mm
{\baselineskip 5mm \tt
\leftline{DO  xyzt=0,NXYZT2-1} 
\leftline{\quad c\%fc(xyzt)=c1\%fc(xyzt)+c2\%fc(xyzt)}
\leftline{END DO}
}

and

\vskip 4mm
{\baselineskip 5mm \tt
\leftline{DO  xyzt=0,NXYZT2-1} 
\leftline{\quad c\%fc(xyzt)=c1\%fc(xyzt)+complex}
\leftline{END DO}
}
\leftline{respectively.}

The following operations are defined and have the obvious meaning, 
implicit in the symbol:

\vskip 1mm

\leftline{$\tt g_1+g_2$,\quad $\tt g_1-g_2$,\quad $\tt g_1*g_2$,
\quad   $\tt g*f$,\quad $\tt f*g$,\quad $\tt g*c$,\quad $\tt c*g$, 
\quad $\tt g/c$,} 
 
\leftline{$\tt g*r$,\quad $\tt r*g$,\quad $\tt g/r$,\quad $\tt g+m$,
\quad $\tt m+g$,\quad $\tt g-m$,\quad $\tt m-g$,\quad $\tt g*m$,
\quad $\tt m*g$,}

\leftline{$\tt g*complex$,\quad $\tt complex*g$,\quad $\tt g/complex$,
\quad $\tt g*real$,\quad $\tt real*g$,\quad $\tt g/real$;}

\vskip 3mm

\leftline{$\tt f_1+f_2$,\quad $\tt f_1-f_2$,\quad $\tt f*c$,\quad $\tt c*f$,
\quad $\tt f/c$,\quad $\tt f*r$,\quad $\tt r*f$,\quad $\tt f/r$,
$\tt \; f*m$, $\tt \; m*f$,}

\leftline{$\tt f*complex$,\quad $\tt complex*f$,\quad $\tt f/complex$,
\quad $\tt f*real$,\quad $\tt real*f$, \quad $\tt f/real$;} 
  
\vskip 3mm

\leftline{$\tt c_1+c_2$,\quad $\tt c_1-c_2$,\quad $\tt c_1*c_2$,
\quad $\tt c_1/c_2$, \quad $\tt c+r$,\quad $\tt r+c$,\quad $\tt c-r$,
\quad $\tt r-c$,}

\leftline{$\tt c*r$,\quad $\tt r*c$,\quad $\tt c/r$,\quad $\tt r/c$,
\quad $\tt c+complex$,\quad $\tt complex+c$,\quad $\tt c-complex$,}

\leftline{$\tt complex-c$,\quad $\tt c*complex$,\quad $\tt complex*c$,
\quad $\tt c/complex$,\quad $\tt complex/c$,}

\leftline{$\tt c+real$,\quad $\tt real+c$,\quad $\tt c-real$,
\quad $\tt real-c$,\quad $\tt c*real$,\quad $\tt real*c$,}

\leftline{$\tt c/real$,\quad $\tt real/c$;}

\vskip 3mm

\leftline{$\tt r_1+r_2$,\quad $\tt r_1-r_2$,\quad $\tt r_1*r_2$,
\quad $\tt r_1/r_2$, \quad $\tt r+real$,\quad $\tt real+r$,}

\leftline{$\tt r-real$,\quad $\tt real-r$,\quad $\tt r*real$,
\quad $\tt real*r$,\quad $\tt r/real$,\quad $\tt real/r$;}

\vskip 3mm

\leftline{$\tt ge_1+ge_2$,\quad $\tt ge_1-ge_2$,\quad $\tt ge*r$,
\quad $\tt r*ge$,\quad $\tt ge/r$,\quad $\tt ge*real$,}

\leftline{$\tt real*ge$,\quad $\tt ge/real$;}

\vskip 3mm

\leftline{$\tt m_1+m_2$,\quad $\tt m_1-m_2$,\quad $\tt m_1*m_2$,
\quad $\tt m*complex$,\quad $\tt complex*m$,\quad $\tt m/complex$,}

\leftline{$\tt m*real$,\quad $\tt real*m$,\quad $\tt m/real$;}

\vskip 2mm

We do not  provide any clarification about the operations
listed above (it would be truly superfluous)
but for the observation that the symbol $\tt *$ implies matrix multiplication
when acting between operands of type gauge\_field or matrix,
and matrix by vector or vector by matrix when one of the operand is 
a fermi\_field and the other a gauge\_field or a matrix. Notice that there
is no implicit complex conjugation of the vector at the r.h.s. of a
vector by matrix multiplication, i.e. $\tt f=f_1 * m$ translates into

\vskip 4mm
{\baselineskip 5mm \tt
\leftline{DO  s=1,4} 
\leftline{DO  xyzt=0,NXYZT2-1} 
\leftline{DO  i=1,3}
\leftline{\quad f\%fc(i,xyzt,s)=f1\%fc(1,xyzt,s)*m\%mc(1,i)\quad \&}
\leftline{\quad \quad \quad +f1\%fc(2,xyzt,s)*m\%mc(2,i)
+f1\%fc(3,xyzt,s)*m\%mc(3,i)}
\leftline{END DO}
\leftline{END DO}
\leftline{END DO}
}

The following additional operations have a special meaning:

\leftline{$\tt g_1/g_2$ :}

{\parindent=0pt
the gauge field $\tt g_1$ is multiplied, site by site, by the Hermitian
adjoint of the gauge field $\tt g_2$ (the notation is motivated by the fact
that, with unitary matrices, the Hermitian adjoint of a matrix is also
its inverse; however, there is no restriction that the variables stored
in a gauge field must represent unitary matrices).}
                                                                
\leftline{$\tt m/g_2$ and $\tt g_1/m$ : same as above, 
but with $\tt m$ a matrix rather than a gauge}
\leftline{field.}

\leftline{$\tt g_1//g_2$ :  the Hermitian adjoint of the gauge 
field $\tt g_1$ is multiplied, site by}
\leftline{site, by the gauge field $\tt g_2$.}
                                                                
\leftline{$\tt g_1//m$ and $\tt m//g_2$ : same as above, 
but with $\tt m$ a matrix rather than a gauge} 
\leftline{field.}

\leftline{$\tt f/g \;$ and $\tt g//f $ : the Fermi field $\tt f$ is
right or left multiplied, site by site, by} 
\leftline{the Hermitian adjoint of the gauge field $\tt g$.} 

\leftline{$\tt f/m \;$ and $\tt m//f $ : same as above, 
but with $\tt m$ a matrix rather than a gauge}
\leftline{field.}

\leftline{$\tt f_1*f_2$ :}  

{\parindent=0pt
this operation returns a complex field having as site components
the scalar product, taken over the color and the spin indices, of
the complex conjugate of $\tt f_1$ and $\tt f_2$. Explicitly,
$\tt c=f_1*f_2$ would be implemented as}
\vskip 4mm
{\baselineskip 5mm \tt
\leftline{DO  xyzt=0,NXYZT2-1} 
\leftline{\quad c\%fc(xyzt)=0}
\leftline{DO  s=1,4} 
\leftline{DO  i=1,3}
\leftline{\quad c\%fc(xyzt)=c\%fc(xyzt) \quad \&}
\leftline{\quad \quad \quad +CONJG(f1\%fc(i,xyzt,s))*f2\%fc(i,xyzt,s)}
\leftline{END DO}
\leftline{END DO}
\leftline{END DO}
}

\leftline{$\tt f_1//f_2$ :}  

{\parindent=0pt
this operation returns a variable of type gauge\_field having as 
site components the dyadic (over the color indices) of
$\tt f_1$ and the complex conjugate of $\tt f_2$. The spin indices
are summed over. Explicitly, $\tt g=f_1//f_2$ would be implemented as}
\vskip 4mm
{\baselineskip 5mm \tt
\leftline{DO  xyzt=0,NXYZT2-1} 
\leftline{DO  i=1,3}
\leftline{DO  j=1,3}
\leftline{\quad g\%fc(i,j,xyzt)=f1\%fc(i,xyzt,1)
*CONJG(f2\%fc(j,xyzt,1))}
\leftline{\quad DO  s=2,4} 
\leftline{\quad \quad g\%fc(i,j,xyzt)=g\%fc(i,j,xyzt) \quad \&}
\leftline{\quad \quad \quad \quad
+f1\%fc(i,xyzt,s)*CONJG(f2\%fc(j,xyzt,s))}
\leftline{\quad END DO}
\leftline{END DO}
\leftline{END DO}
\leftline{END DO}
}

\leftline{$\tt ge_1*ge_2$ : this operation returns a real field having 
as site components the}
\leftline{scalar product of the site components of the generators.}

\leftline{$\tt g_1.Dot.g_2$  : this operation returns a real field having 
as site components the} 
\leftline{the real part of the trace of the product of the Hermitian
adjoint of the site} 
\leftline{components of the gauge field $\tt g_1$ with the site components
of the gauge} 
\leftline{field $\tt g_2$.}

The following named operators are also defined:

{\parindent=0pt
$\tt i.Gamma.f$ , where $\tt i$ is a scalar integer.  
This operation returns a Fermi field
having as site components the product of a single $\gamma$ matrix
or of a pair of $\gamma$ matrices times the site components $\psi_{\bf x}$
of the Fermi field $\tt f$.  Our convention is as follows. The integer
variable $\tt i$ can take value $1$ through $5$ or value
$10*i_1+i_2$, where $i_1$ and $i_2$ can again range from
$1$ to $5$.  In the former case the operator implements the product
$\gamma_i \psi_{\bf x}$. In the latter case the pair $i_1,i_2$ 
stands for two indices labeling a matrix 
$\gamma_{i{\scriptscriptstyle1}\,i{\scriptscriptstyle2}}$, where
$\gamma_{i{\scriptscriptstyle1}\,i{\scriptscriptstyle2}}=
{\imath \over 2} [\gamma_{i{\scriptscriptstyle1}}
\gamma_{i{\scriptscriptstyle2}}-
\gamma_{i{\scriptscriptstyle2}}
\gamma_{i{\scriptscriptstyle1}}]$,  
$\gamma_{i\,5}=-\gamma_{5\,i}=\gamma_i \gamma_5$ with
$i,i_1,i_2 = 1 \dots 4$, and the operator implements the product
$\gamma_{i{\scriptscriptstyle1}\,i{\scriptscriptstyle2}}\psi_{\bf x}$.
Thus, for instance, $\tt i=25; f1=i.Gamma.f2$ would implement
$\psi_{1{\bf x}} = \gamma_2 \gamma_5 \psi_{2{\bf x}}$.  Products
of $\gamma$ matrices have been explicitly incorporated in the
definition of the $\tt .Gamma.$ operator because they are
frequently encountered in the evaluation of matrix elements
of fermionic variables.

$\tt f.Gamma.i$ , where $\tt i$ is a scalar integer.  
This operation returns a Fermi field
having as site components the product of site components 
of the Fermi field $\tt f$ times a single $\gamma$ matrix
or of a pair of $\gamma$ matrices, following the same convention
about the values of $\tt i$ as above.

$\tt i.Lambda.g$ , where $\tt i$ is a scalar integer.  
This operation returns a gauge field
having as site components the product of the matrix $\lambda_i$ 
times the site components of the gauge field $\tt g$.

$\tt g.Lambda.i$ , where $\tt i$ is a scalar integer. 
This operation returns a gauge field
having as site components the product of the site components 
of the gauge field $\tt g$ times the matrix $\lambda_i$.}

In addition we define the following unary operators:

\centerline{$\tt .I.$,\quad $\tt .Minus.$,\quad $\tt .Conjg.$,
\quad $\tt .Adj.$, \quad $\tt .Ctr.$  \quad $\tt .Tr.$ \quad
$\tt .Sqrt.$ \quad and \quad $\tt .Exp.$}

When acting on a variable of type gauge\_field, fermi\_field or 
complex\_field $\tt .I.$ returns $\imath$ times the variable.  When
acting on a variable of type real\_field it returns a complex
field given by $\imath$ times the real field.  This is introduced
for efficiency, since the operator is implemented by switching
real and imaginary parts with the appropriate change of sign, rather than
through a complex multiplication.
 
When acting on a variable of type gauge\_field, fermi\_field, 
complex\_field, real\_field or generator\_field,  
$\tt .Minus.$ returns the negative of the variable.
 
When acting on a variable of type gauge\_field, fermi\_field, 
complex\_field or matrix $\tt .Conjg.$ returns the complex conjugate
of the variable, i.e. a variable whose complex components are the
complex conjugate of the original one. 
 
When acting on a variable of type gauge\_field or matrix $\tt .Adj.$ 
returns the Hermitian adjoint of the variable.

When acting on a variable of type gauge\_field or matrix $\tt .Ctr.$ 
returns a complex\_field or complex number, respectively,
equal to the trace (at each site in the case of a field) of
the operand.

When acting on a variable of type gauge\_field or matrix $\tt .Tr.$ 
returns a real\_field or real number, respectively,
equal to the real part of the trace (at each site in the case of a field) 
of the operand.
 
When acting on a variable of type real\_field $\tt .Sqrt.$ 
returns a real\_field having as site components the square root
of the absolute value of the site components of the operand.
At the same time the global variable context is set to $\tt .TRUE.$
at all sites where the operand is non-negative and to $\tt .FALSE.$
at all other sites.
 
When acting on a variable of type real\_field $\tt .Exp.$ 
returns a real\_field having as site components the exponential
of the site components of the operand.

\subsection{Modules assign\_isotype1, assign\_isotype2, \newline 
assign\_isotype3 and assign\_mixed}
\label{assign}

These modules define the normal assignment and a variety of
overloaded assignments which are defined for efficiency 
(cf.~Sect.~\ref{overloaded} above) and programming convenience.
They are presented as four separate modules (assign\_isotype1,
assign\_isotype2  and assign\_isotype3 define assignments 
between variables of the same type,
assign\_mixed between variables of different type) to
reduce the overall length of the individual modules.  
We reproduce here all the
available assignments.  We use the notational conventions
we introduced at the beginning of Sect.~\ref{fieldalgebra}.
Namely, we use the symbols $\tt g,$ $\tt u,$
$\tt f,$ $\tt c,$ $\tt r,$ $\tt ge$ and $\tt m$ 
to denote variables of type  gauge\_field, full\_gauge\_field,
fermi\_field,  complex\_field, real\_field, 
generator\_field and  matrix, respectively, and the
symbols $\tt complex$ and $\tt real$ to denote a complex or real variable
of kind $\tt REAL8$ (cf.~Sect.~\ref{precisions}).  
Also, we use subscripts, e.g.~$\tt f_1, f_2$, to distinguish between two 
variables of the same type.

When the assignment relies on the the fact that the global 
variables $\tt assign\_type$ 
and $\tt assign\_spec$ have a value different from their default
values $'='$ and $0$, we will denote this fact by the
using the combined symbols $\tt assign\_type (assign\_spec) = $
to denote the assignment.  For example, we would use $\tt f_1 \; += f_2 $ or
$\tt f_1 \; U(2)= f_2 $ to denote assignments which in the actual coding
would be implemented as

\leftline{{\tt assign\_type='+'; f1=f2} , or}

\leftline{{\tt assign\_type='U'; assign\_spec=2; f1=f2} , respectively.} 

A general rule is that all assignments set the global
variables  $\tt assign\_type$ and $\tt assign\_spec$ 
equal to their default  values $'='$ and $0$, 
no matter what the assignment does.  As discussed
in Sect.~\ref{overloaded}, this is done in order 
to avoid the accidental use of erroneous assignments.

For the $\tt parity$ component, 
the rule is that, if the destination is not an operand in the assignment 
(i.e.~it is a variable with strict $\tt INTENT(OUT)$),
the $\tt parity$ component (if present) 
of the variable at the l.h.s.~of the assignment (destination) 
is set equal to the $\tt parity$ of the variable at the 
r.h.s.~of the assignment (source), or set to $\tt -1$ if the
source has no $\tt parity$.
Similarly, when the destination is not an operand in the
assignment and has a $\tt dir$ component, this is set equal
to the  $\tt dir$ of the source or to $\tt 0$
if the source has no $\tt dir$. 
An exception to the rule above about the $\tt parity$ component occurs with 
the $\tt assign\_type = 'u'$, $\tt assign\_type = 'w'$ 
and $\tt assign\_type = 'x'$ assignments, 
which copy into the destination a shifted source.  In this case, 
if the $\tt parity$ of the source is defined, the $\tt parity$ of 
the destination is set to the opposite value.

If the destination is an operand in
the assignment (i.e.~it is a variable with $\tt INTENT(INOUT)$)
$\tt parity$ and $\tt dir$ are treated in a manner similar to
what happens in the definition of the operators implemented by the
overloaded assignment.  Typically, if the destination and the
other operand have the same $\tt parity$, this is preserved, otherwise
the $\tt parity$ of the destination is set to $\tt -1$ (undefined).
An exception is found in the assignments $\tt U=$ and $\tt W=$
which implement the sum of the destination with a shifted operand,
in which case the $\tt parity$ of the destination is preserved
if the other operand has the opposite $\tt parity$ (as is the
case in a geometrically meaningful operation) and is returned
undefined otherwise.

In what follows we list all of the available assignments 
and define their action,
appending a few words of explanation when appropriate.  
When the assignment is not followed by further clarifications,
it means that it is a straightforward assignment 
(with $\tt assign\_type \; '='$) copying  
the content of the source into the destination.  Also, whenever the 
assignment implements operations which can be performed by using
overloaded operators, we illustrate its action
simply by reformulating it in terms of these operators.  We refer to the
sections detailing the modules where the overloaded operators
are defined for clarification of their action.

The assignments are listed in order of destination type, first,
and then of source type. The ordering of the types is the same as
their order of introduction in Sect.~\ref{types}. 

Available assignments:

\leftline{$\tt g_1=g_2$} 

\leftline{$\tt g_1\;+=g_2$ \quad \quad ($\tt g_1=g_1+g_2$)}

\leftline{$\tt g_1\;-=g_2$ \quad \quad  ($\tt g_1=g_1-g_2$)}

\leftline{$\tt g_1\;*(0)=g_2$ \quad \quad  ($\tt g_1=g_1*g_2$)}

\leftline{$\tt g_1\;*(-1)=g_2$ \quad \quad  ($\tt g_1=g_2*g_1$)}

Notice how the $\tt assign\_spec$ variable is used, above and
immediately below, to specify the
order of the operands in the non-commutative matrix multiplication.

\leftline{$\tt g_1\;/(0)=g_2$ \quad \quad  ($\tt g_1=g_1/g_2$)}

\leftline{$\tt g_1\;/(-1)=g_2$ \quad \quad  ($\tt g_1=g_2//g_1$)}

\leftline{$\tt g_1\;u(dir)=g_2$ \quad \quad ($\tt g_1=dir.Ushift.g_2$)}

\leftline{$\tt g_1\;U(dir)=g_2$ \quad \quad ($\tt g_1=g_1+(dir.Ushift.g_2)$)}

\leftline{$\tt g_1\;t=g_2$ \quad \quad  ($\tt g_1=g_2$ where $\tt context$
is $\tt .TRUE.$)}

\leftline{$\tt g_1\;f=g_2$ \quad \quad  ($\tt g_1=g_2$ where $\tt context$
is $\tt .FALSE.$)}

\leftline{$\tt g_1\;A=g_2$ \quad \quad ($\tt g_1=.Adj.g_2$)} 

\leftline{$\tt g_1\;C=g_2$ \quad \quad ($\tt g_1=.Conjg.g_2$)} 

\leftline{$\tt g_1\;I=g_2$ \quad \quad ($\tt g_1=.I.g_2$)} 

\leftline{$\tt g_1\;M=g_2$ \quad \quad ($\tt g_1=.Minus.g_2$)} 

\leftline{$\tt g=u$ \quad \quad  ($\tt u\%uc(g\%parity,g\%dir)$ is copied to 
$\tt g\%fc$)}

\leftline{$\tt g\;t=u$ \quad \quad (same as above, but only 
where $\tt context$ is $\tt .TRUE.$)}

\leftline{$\tt g\;f=u$ \quad \quad (same as two lines above, but only 
where $\tt context$ is $\tt .FALSE.$)}

\leftline{$\tt g=ge$ \quad \quad ($\tt g=.Matrix.ge$)} 

\leftline{$\tt g\;E=ge$ \quad \quad ($\tt g=.Exp.ge$)} 

\leftline{$\tt g=m$ \quad \quad (all elements of g are set equal to $\tt m$)}

\leftline{$\tt g\;*(0)=m$ \quad \quad ($\tt g=g*m$)}

\leftline{$\tt g\;*(-1)=m$ \quad \quad ($\tt g=m*g$)}

\leftline{$\tt g\;*=complex$ \quad \quad ($\tt g=g*complex$)}

\leftline{$\tt g\;*=real$ \quad \quad ($\tt g=g*real$)}

\leftline{$\tt g\;/=complex$ \quad \quad ($\tt g=g/complex$)}

\leftline{$\tt g\;/=real$ \quad \quad ($\tt g=g/real$)}

\leftline{$\tt u=g$ \quad \quad ($\tt g\%fc$ is copied to 
$\tt u\%uc(g\%parity,g\%dir)$ )}

\leftline{$\tt u\;t=g$ \quad \quad (same as above, but only 
where $\tt context$ is $\tt .TRUE.$)}

\leftline{$\tt u\;f=g$ \quad \quad (same as two lines above, but only 
where $\tt context$ is $\tt .FALSE.$)}

\leftline{$\tt u_1=u_2$}

\leftline{$\tt f\;*(0)=g$\quad \quad ($\tt f=f*g$)}  

\leftline{$\tt f\;*(-1)=g$\quad \quad ($\tt f=g*f$)}  

\leftline{$\tt f\;/(0)=g$\quad \quad ($\tt f=f/g$)} 

\leftline{$\tt f\;/(-1)=g$\quad \quad ($\tt f=g//f$)} 

(Note the function played by $\tt assign\_spec$ in the four preceding 
assignments.)

\leftline{$\tt f_1=f_2$} 

\leftline{$\tt f_1\;+=f_2$ \quad \quad ($\tt f_1=f_1+f_2$)} 

\leftline{$\tt f_1\;-=f_2$ \quad \quad ($\tt f_1=f_1-f_2$)} 

\leftline{$\tt f_1\;u(dir)=f_2$ \quad \quad ($\tt f_1=dir.Ushift.f_2$)} 

\leftline{$\tt f_1\;U(dir)=f_2$ \quad \quad ($\tt f_1=f_1+(dir.Ushift.f_2)$)} 

\leftline{$\tt f_1\;w(dir)=f_2$ \quad \quad ($\tt f_1=dir.Wshift.f_2$)} 

\leftline{$\tt f_1\;W(dir)=f_2$ \quad \quad ($\tt f_1=f_1+(dir.Wshift.f_2)$)} 

\leftline{$\tt f_1\;x(dir)=f_2$ \quad \quad ($\tt f_1=dir.Xshift.f_2$)} 

\leftline{$\tt f_1\;X(dir)=f_2$ \quad \quad ($\tt f_1=f_1+(dir.Xshift.f_2)$)} 

\leftline{$\tt f_1\;C=f_2$ \quad \quad ($\tt f_1=.Conjg.f_2$)} 

\leftline{$\tt f_1\;I=f_2$ \quad \quad ($\tt f_1=.I.f_2$)} 

\leftline{$\tt f_1\;M=f_2$ \quad \quad ($\tt f_1=.Minus.f_2$)} 

\leftline{$\tt f\;*=c$ \quad \quad ($\tt f=f*c$)} 

\leftline{$\tt f\;/=c$ \quad \quad ($\tt f=f/c$)} 

\leftline{$\tt f\;*=r$ \quad \quad ($\tt f=f*r$)} 

\leftline{$\tt f\;/=r$ \quad \quad ($\tt f=f/r$)} 

\leftline{$\tt f\;*=complex$ \quad \quad ($\tt f=f*complex$)} 

\leftline{$\tt f\;*=real$ \quad \quad ($\tt f=f*real$)} 

\leftline{$\tt f\;/=complex$ \quad \quad ($\tt f=f/complex$)} 

\leftline{$\tt f\;/=real$ \quad \quad ($\tt f=f/real$)} 

\leftline{$\tt c=g$ \quad \quad ($\tt c=.Ctr.g$)} 

\leftline{$\tt c_1=c_2$} 

\leftline{$\tt c_1\;+=c_2$ \quad \quad ($\tt c_1=c_1+c_2$)} 

\leftline{$\tt c_1\;-=c_2$ \quad \quad ($\tt c_1=c_1-c_2$)} 

\leftline{$\tt c_1\;*=c_2$ \quad \quad ($\tt c_1=c_1*c_2$)} 

\leftline{$\tt c_1\;/=c_2$ \quad \quad ($\tt c_1=c_1/c_2$)} 

\leftline{$\tt c_1\;C=c_2$ \quad \quad ($\tt c_1=.Conjg.c_2$)} 

\leftline{$\tt c_1\;M=c_2$ \quad \quad ($\tt c_1=.Minus.c_2$)} 

\leftline{$\tt c_1\;I=c_2$ \quad \quad ($\tt c_1=.I.c_2$)} 

\leftline{$\tt c=r$} 

\leftline{$\tt c\;+=r$ \quad \quad ($\tt c=c+r$)} 

\leftline{$\tt c\;-=r$ \quad \quad ($\tt c=c-r$)}  

\leftline{$\tt c\;*=r$ \quad \quad ($\tt c=c*r$)} 

\leftline{$\tt c\;/=r$ \quad \quad ($\tt c=c/r$)} 

\leftline{$\tt c\;M=r$ \quad \quad ($\tt c=.Minus.r$)} 

\leftline{$\tt c=complex$} 

\leftline{$\tt c\;+=complex$ \quad \quad ($\tt c=c+complex$)} 

\leftline{$\tt c\;-=complex$ \quad \quad ($\tt c=c-complex$)}  

\leftline{$\tt c\;*=complex$ \quad \quad ($\tt c=c*complex$)} 

\leftline{$\tt c\;/=complex$ \quad \quad ($\tt c=c/complex$)} 

\leftline{$\tt c\;M=complex$ \quad \quad ($\tt c=.Minus.complex$)} 

\leftline{$\tt c=real$} 

\leftline{$\tt c\;+=real$ \quad \quad ($\tt c=c+real$)} 

\leftline{$\tt c\;-=real$ \quad \quad ($\tt c=c-real$)}  

\leftline{$\tt c\;*=real$ \quad \quad ($\tt c=c*real$)} 

\leftline{$\tt c\;/=real$ \quad \quad ($\tt c=c/real$)} 

\leftline{$\tt c\;M=real$ \quad \quad ($\tt c=.Minus.real$)} 

\leftline{$\tt r=g$ \quad \quad ($\tt r=.Tr.g$)} 

\leftline{$\tt r=f$ \quad \quad ($\tt r=f*f$)} 

\leftline{$\tt r=c$ \quad \quad the elements of $\tt r$ are set equal to the
real part of the elements of $\tt c$} 

\leftline{$\tt r_1=r_2$} 

\leftline{$\tt r_1\;+=r_2$ \quad \quad ($\tt r_1=r_1+r_2$)} 

\leftline{$\tt r_1\;-=r_2$ \quad \quad ($\tt r_1=r_1-r_2$)} 

\leftline{$\tt r_1\;*=r_2$ \quad \quad ($\tt r_1=r_1*r_2$)} 

\leftline{$\tt r_1\;/=r_2$ \quad \quad ($\tt r_1=r_1/r_2$)} 

\leftline{$\tt r_1\;M=r_2$ \quad \quad ($\tt r_1=.Minus.r_2$)} 

\leftline{$\tt r_1\;R=r_2$ \quad \quad ($\tt r_1=.Sqrt.r_2$)} 

\leftline{$\tt r_1\;E=r_2$ \quad \quad ($\tt r_1=.Exp.r_2$)} 

\leftline{$\tt r=real$} 

\leftline{$\tt r\;+=real$ \quad \quad ($\tt r=r+real$)} 

\leftline{$\tt r\;-=real$ \quad \quad ($\tt r=r-real$)} 

\leftline{$\tt r\;*=real$ \quad \quad ($\tt r=r*real$)} 

\leftline{$\tt r\;/=real$ \quad \quad ($\tt r=r/real$)} 

\leftline{$\tt r\;M=real$ \quad \quad ($\tt r=.Minus.real$)} 

\leftline{$\tt ge=g$ \quad \quad ($\tt ge=.Generator.g $)}  

\leftline{$\tt ge_1=ge_2$} 

\leftline{$\tt ge_1\;+=ge_2$ \quad \quad ($\tt ge_1=ge_1+ge_2$)}  

\leftline{$\tt ge_1\;-=ge_2$ \quad \quad ($\tt ge_1=ge_1-ge_2$)}   

\leftline{$\tt ge_1\;M=ge_2$ \quad \quad ($\tt ge_1=.Minus.ge_2$)}

\leftline{$\tt ge_1\;S=ge_2$ \quad \quad ($\tt ge_1=.Sq.ge_2$)}

\leftline{$\tt ge\;*=r$ \quad \quad ($\tt ge=ge*r$)}

\leftline{$\tt ge\;/=r$ \quad \quad ($\tt ge=ge/r$)}

\leftline{$\tt ge\;*=real$ \quad \quad ($\tt ge=ge*real$)}

\leftline{$\tt ge\;/=real$ \quad \quad ($\tt ge=ge/real$)}

The following assignments perform global reductions, either absolute
or restricted to the lattice sites where $\tt context$ is $\tt .TRUE.$
or $\tt .FALSE.$:   

\leftline{$\tt complex=c$ \quad \quad 
($\tt complex=\sum_{xyzt} c(xyzt)$)}   

\leftline{$\tt complex\;t=c$ \quad \quad 
($\tt complex=\sum_{WHERE(context(xyzt))} c(xyzt)$)}   

\leftline{$\tt complex\;f=c$ \quad \quad 
($\tt complex=\sum_{WHERE(.NOT.context(xyzt))} c(xyzt)$)}   

\leftline{$\tt real=c$ \quad \quad 
($\tt real=\sum_{xyzt} Real[c(xyzt)]$)}   

\leftline{$\tt real\;t=c$ \quad \quad 
($\tt real=\sum_{WHERE(context(xyzt))} Real[c(xyzt)]$)}   

\leftline{$\tt real\;f=c$ \quad \quad 
($\tt real=\sum_{WHERE(.NOT.context(xyzt))} Real[c(xyzt)]$)}   

\leftline{$\tt real=r$ \quad \quad 
($\tt real=\sum_{xyzt} r(xyzt)$)}   

\leftline{$\tt real\;t=r$ \quad \quad 
($\tt real=\sum_{WHERE(context(xyzt))} r(xyzt)$)}   

\leftline{$\tt real\;f=r$ \quad \quad 
($\tt real=\sum_{WHERE(.NOT.context(xyzt))} r(xyzt)$)}   

\subsection{Module shifts}
\label{shifts}

This module defines the operators $\tt .Cshift.$, $\tt .Ushift.$,
$\tt .Wshift.$ and $\tt .Xshift.$ The left operand for all these 
operators is an integer $\tt m$ which must take one of the values 
$\tt 1,2,3,4,-1,-2,-3,-4$ and specifies the direction and 
orientation of the shift.  The right operand can be any variable of
field type for $\tt .Cshift.$. It can be any variable of field type with the
exception of the type generator\_field for $\tt .Ushift.$, while it must be
a variable of type fermi\_field for $\tt .Wshift.$ and $\tt .Xshift.$  
The $\tt parity$
component of the right operand must be defined, i.e.~take value $\tt 0$
or $\tt 1$.  If the $\tt parity$ is not defined or if $\tt m$ does
not take one of the values specified above the function call 
implementing the operator returns an error message and stops the program.
All of these operators return a field variable of the same type as
the left operand and opposite $\tt parity$.  If the right operand
is of the type gauge\_field or generator\_field and thus has also
a $\tt dir$ component, this is passed to the result unchanged.

$\tt .Cshift.$ implements an ordinary C-shift of the field, but with
respect to the Cartesian geometry of the lattice.  This is why the
parity is interchanged.  Given a site with Cartesian coordinates
$\bf x$ in the sublattice of the parity of the result, the operator
copies into the corresponding element of the result the element of the 
right operand which is defined over the lattice site ${\bf x} +s \hat\mu$,
where $s$ is the sign of $\tt m$ and $\mu = {\rm abs}({\tt m})$.

$\tt .Ushift.$ moves the data in a manner similar to $\tt .Cshift.$,
but with the inclusion of the appropriate transport factors, defined
in terms of the global field $\tt U$ (cf.~``global module''). 
For the variable of type gauge\_field and fermi\_field the U-shift
operation has been defined in Sect.~\ref{thenotion} (cf.~Eqs.~(\ref{shiftv}),
(\ref{negshiftv}), and ~Eqs.~(\ref{shiftfm}), (\ref{negshiftfm}),
for gauge fields and Fermi fields, respectively).  
The action of a U-shift on variables
of type complex\_field and real\_field reduces to a C-shift.
The U-shift of a generator field is not normally encountered in
QCD simulations and for this reason it is not explicitly implemented here.
It can be implemented by using the functionality provided by the
module generator\_algebra to re-express the generator field as a field
of hermitian matrices (i.e.~of type gauge\_field), shifting the latter,
and converting it again to a generator\_field.

The operator $\tt .Wshift.$ acts only on Fermi fields and it is a 
combination of a $\tt .Ushift.$ and the product with a $\gamma$
matrix.  Precisely, if we again define $s$ to be the sign of $\tt m$
and define $\mu$ or $\tt mu$ to be the absolute value of $\tt m$,
then the operation $\tt f2=m.Wshift.f1$ is equivalent to
 
\leftline{\tt f2=m.Ushift.f1-s*(mu.Gamma.(m.Ushift.f1))}

Equivalently
\begin{equation}
f_{2,{\bf x}}= (1-\gamma_{\mu})U^{\mu}_{{\bf x}} 
f_{1,{\bf x}+\hat\mu}
\label{pwilson}
\end{equation}
for positive $\tt m$, and
\begin{equation}
f_{2,{\bf x}}= (1+\gamma_{\mu})
U^{\dagger \mu}_{{\bf x}-\hat\mu} f_{1,{\bf x}-\hat\mu}
\label{mwilson}
\end{equation}
for negative $\tt m$.

The operator $\tt .Xshift.$ also acts only on Fermi fields and it is 
equivalent to a W-shift bracketed by two matrices $\gamma_5$
(where $\gamma_5=\gamma_1 \gamma_2 \gamma_3 \gamma_4$):

\begin{equation}
f_{2,{\bf x}}= \gamma_5 (1-\gamma_{\mu}) \gamma_5 U^{\mu}_{{\bf x}} 
f_{1,{\bf x}+\hat\mu}
\label{pxwilson}
\end{equation}
for positive $\tt m$, and
\begin{equation}
f_{2,{\bf x}}= \gamma_5 (1+\gamma_{\mu}) \gamma_5
U^{\dagger \mu}_{{\bf x}-\hat\mu} f_{1,{\bf x}-\hat\mu}
\label{mxwilson}
\end{equation}
for negative $\tt m$.

The reason for the explicit introduction of the W-shift and X-shift
operators is that the combinations (\ref{pwilson}-\ref{mxwilson}) 
appear in the Wilson
discretization of the  Dirac operator, which is a very
widely used lattice Dirac operator, and related algebra,
and that the calculation of the l.h.s.~of Eqs.~(\ref{pwilson}-\ref{mxwilson}) 
is one of the most
time consuming tasks of any QCD simulation.  Moreover, the combinations
$1 \pm \gamma_{\mu}$ appearing in (\ref{pwilson}-\ref{mxwilson}) 
are projection operators, which effectively limits the U-shift to
a subspace of the spin space of dimensionality two.  Thus a direct
implementation of the W-shift, rather than via a combination of
the $\tt .Ushift.$ and $\tt .Gamma.$ operators, entails substantial
advantages of efficiency.

\subsection{Module Dirac\_operator}     
\label{diracop}

The (Wilson) lattice Dirac operator, acting on a Fermi field 
$f_1$, produces a Fermi field 
$f_2$, given by 
\begin{equation}
f_{2,{\bf x}}= \sum_{\mu} [(1-\gamma_{\mu})U^{\mu}_{{\bf x}} 
f_{1,{\bf x}+\hat\mu} + (1+\gamma_{\mu})
U^{\dagger \mu}_{{\bf x}-\hat\mu} f_{1,{\bf x}-\hat\mu}]
\label{dirac}
\end{equation}
It is obvious from this equation that the lattice Dirac operator 
only connects components of Fermi fields of opposite parity.
The unary operator $\tt .Dirac.$ accepts as operand a variable of
type fermi\_field, which must have a definite $\tt parity$,
and returns a variable of the same type and opposite $\tt parity$
given by the action of the lattice Dirac operator~(\ref{dirac})
on the operand.  The unary operator $\tt .XDirac.$ implements the
action of the Dirac operators bracketed by two matrices $\gamma_5$,
i.e.~, if $\tt f$ is a variable of type fermi\_field,  
$\tt .XDirac.f$ returns the same results as 
$\tt i5.Gamma.(.XDirac.(i5.Gamma.f))$, where the integer variable
$\tt i5$ equals 5.

In this module the operators $\tt .Dirac.$ and $\tt .XDirac.$ are  
implemented using the operators $\tt .Wshift.$ and $\tt .Xshift.$,  
introduced in the module field\_algebra.  We have defined them 
as separate operators for convenience of coding and
also because, the application of these operators being the most
CPU intensive part for the majority of applications, this module
isolates the code whose optimization would produce the largest
returns.  A programmer striving for exceptional efficiency might
want to code this module as a highly optimized, self-standing
implementation of the lattice Dirac operator.  Even if this route
is chosen, we are certain that the advantages of having a module
written at a higher level against which to compare the results
of the optimized module are not lost on the practicing 
computational scientist.

\subsection{Module generator\_algebra}  
\label{generatoralgebra}

This module defines the unary operators $\tt .Matrix.$, $\tt .Generator.$,
$\tt .Sq. $ and $\tt .Exp.$ which perform some special operations
involving generator fields.  The operators accept arguments of the
type generator\_field or gauge\_field and return as result a variable
of one of these types.  The $\tt parity$ and $\tt dir$ components of the
argument are passed on to the result.

$\tt .Matrix.$ accepts an argument $\tt ge$ of type generator\_field
and returns a result $\tt v$ of type gauge\_field containing, site by site, 
the Hermitian matrix
\begin{equation}
v_{ij,{\bf x}} =\sum_k \lambda^k_{ij}\, ge_{k,{\bf x}} \ .
\label{matrix}
\end{equation}

$\tt .Generator.$ accepts an argument  $\tt v$ of the type gauge\_field
and returns a result $\tt ge$ of type generator\_field containing, 
site by site, the traceless, antihermitian part of the argument: 
\begin{equation}
ge_{k,{\bf x}} = -{\imath \over 4} \sum_{ij}\lambda^k_{ij}\, (v_{ji,\bf x}
-v^*_{ij,\bf x}) \ .
\label{gen}
\end{equation}
Notice that $\tt .Generator.$ and $\tt .Matrix.$ are not inverse
operators.  However $\tt .Generator.(.Matrix.(IU*ge))$ does return $\tt ge$,
while $\tt .Matrix.(.Generator.v)$ returns the traceless, 
antihermitian part of $\tt v$: $v_{\scriptscriptstyle{AH}}=
(v-v^{\dagger}) / (2 \imath)$.

$\tt .Sq.$ accepts an argument of type generator\_field $\tt ge_1$ 
and returns a result $\tt ge_2$ of the same type containing, site by site, 
the generator corresponding to the traceless part of the square 
of $\tt .Matrix.ge1$: 
\begin{equation}
(ge_2)_{l,{\bf x}} = 
{ 1 \over 2} \sum_{ijk} \big[ \lambda^l_{ij} \,
\big( \sum_m \lambda^m_{jk}\, (ge_1)_{ m,{\bf x}} \big) \,
\big( \sum_n \lambda^n_{ki}\, (ge_1)_{ n,{\bf x}} \big) 
\big]\ .
\label{sq}
\end{equation}

$\tt .Exp.$ accepts an argument $\tt ge$ of type generator\_field
and returns a result $\tt v$ of type gauge\_field containing, site by site, 
the exponentiated generator component:
\begin{equation}
v_{ij,{\bf x}} =\big[ \exp \big(\imath \sum_k \lambda^k\, ge_{k,{\bf x}}\big)
\big]_{ij} \ .
\label{exp}
\end{equation}
The algorithm used for this exponentiation deserves a few words of explanation.
Let us define $H=\sum_k \lambda^k_{ij}\, ge_{k,{\bf x}}$, $q={\rm Tr} H^2$
and $p={\rm Det} H = ({\rm Tr} H^3)/3$.  From the characteristic
equation (recall that ${\rm Tr} H =0$)
\begin{equation}
H^3 - {q \over 2} H - p I =0 \ ,
\label{charact}
\end{equation}
$I$ being the identity matrix, satisfied by $H$ and 
therefore by its eigenvalues $h_n$, we can
easily calculate the eigenvalues as
\begin{equation}
h_n =a \cos[\alpha + 2 \pi (n-1)/3], \quad n=1,2,3 \ ,
\label{eigenval}
\end{equation}
with $a=\sqrt{2 q / 3}$,$\; \alpha = [\cos^{-1}(4 p / a^3)]/3$.  
We order the eigenvalues so that $|h_1| \ge |h_2| \ge |h_3|$.

In a basis where $H$ is diagonal, it is easy to express it as
a linear combination  of two matrices of type ``$\lambda^3\,$'' 
(one 0 eigenvalue) and ``$\lambda^8\,$'' (two degenerate eigenvalues), 
respectively. Of the six different ways in which this can be done 
we use the decomposition
\begin{equation}
H = S + K
\label{spk}
\end{equation}
with $S = {\rm diag} (-h_1-2 h_2, h_1+2 h_2, 0)$, 
$K = {\rm diag} (2(h_1+ h_2), -h_1-h_2, -h_1-h_2)$.

By using the eigenvalues determined above, it is
straightforward to express $S$ in the form 
\begin{equation}
S = c_1 H + c_2 (H^2 - q I /3) \ .
\label{scc}
\end{equation}
Through their dependence on the
eigenvalues and Eq.~(\ref{eigenval}), however, $c_1$ and $c_2$ are
functions of the invariants $q$ and $p$ only.  It follows that 
Eqs.~(\ref{scc}) and (\ref{spk}) provide a decomposition into 
two matrices of type ``$\lambda^3\,$'' and ``$\lambda^8\,$'' irrespective
of the basis.  On the other hand, with a matrix
of type ``$\lambda^3\,$'' it is straightforward to calculate
$\exp(\imath S)$ expressing it as a linear combination
of $I$, $S$ and $S^2$.  Similarly $\exp(\imath K)$ can be expressed
as a linear combination of $I$ and $K$. $\exp (\imath H)$ can be
finally calculated as product of the two commuting
matrices $\exp (\imath S)$ and $\exp (\imath K)$.

It is very important to have an efficient algorithm for the
exponentiation of a matrix, since this operation can be a time
consuming component of several QCD calculations. The algorithm
outlined above has been implemented in the module
``generator\_algebra'' by performing a substantial amount of the
algebra directly in terms of generator components and inlining all of
the operations.  The exponentiation can thus be done with a
reasonably contained number of arithmetic operations, in particular
approximately 300 explicit real multiplications. By way of
comparison, just one product of $3 \times 3$ complex matrices requires
108 real multiplications (i.e.~27 complex multiplications -- these could
also be performed with 81 real multiplications, but then with a much larger
number of adds).

\subsection{Module random\_numbers}
\label{randomnumbers}

This module implements a parallelizable version of the unix pseudorandom
number generator erand48 and provides added functionality.

erand48 is a congruential pseudorandom number generator based on the
iterative formula
\begin{equation}
s_{i+1}=a_1*s_i+b_1 \quad {\rm mod} \; m \ ,
\label{erand}
\end{equation}
where $a_1=\tt 0x5DEECE66D$, $b_1=\tt 0xB$, $m=2^{48}$, $s_i$ 
and $s_{i+1}$ are integers of at least 48 bits of precision.  
The ``seeds'' $s_i$ are converted
to real pseudorandom numbers $r_i$ with uniform distribution 
between $0$ and $1$ by $r_i=2^{-48}\, s_i$.

As presented above, the algorithm is intrinsically serial. However it
follows from Eq.~(\ref{erand}) that the $\rm N^{th}$ iterate $s_{i+N}$
is still of the form $s_{i+N}=a_N*s_i+b_N \; {\rm mod} \; 2^{48}$ 
with integers
$a_N$ and $b_N$ which are uniquely determined by $a_1$, $b_1$.
The module takes advantage of this fact and of the existence of
the global variable $\tt seeds$ (cf.~global\_module) to generate 
pseudorandom numbers in a parallelizable fashion.  

The module defines the following unary operators: $\tt .Seed.$,
$\tt .Rand.$, $\tt .Gauss.$ and $\tt .Ggauss.$.

$\tt .Seed.$ must be used to initialize the generation of pseudorandom
numbers.  When invoked with an argument $\tt saved\_seed$ of kind
$\tt LONG$ (8-byte integer, defined in the module precisions)
$\tt .Seed.$ sets the global variable $\tt seeds$ to the 
sequence~(\ref{erand})
beginning with $\tt saved\_seed$ and also sets the global variables 
$\tt seed\_a$, $\tt seed\_b$ to the appropriate multiplier and constant 
term, $a_N$ and $b_N$, needed to produce increments by $N=NX*NY*NZ*NT/2$ 
in the sequence of pseudorandom numbers.  It returns $\tt .TRUE.$.

When acting on a logical variable equal to $\tt .TRUE.$, $\tt .Seed.$
returns the current seed (=$\tt seeds(0,0,0,0)$), which must
be used to restart the sequence of pseudorandom numbers.
If the argument is $\tt .FALSE.$, $\tt .Seed.$  returns $0$.  

The unary operator $\tt .Rand.$, if invoked with a real argument $\tt real$
of kind $\tt REAL8$, returns a real\_field of pseudorandom numbers 
with uniform distribution between 0 and $\tt real$.  At the same time
it upgrades the global variables $\tt seeds$ using the multiplier $a_N$
(i.e.~$\tt seed\_a$) and constant term $b_N$ (i.e.~$\tt seed\_b$).
It follows that subsequent calls to $\tt .Rand.$ produce real fields  
with the same distribution of pseudorandom numbers which one would
have obtained invoking erand48 sequentially within nested DO loops:
\vskip 4mm
{\baselineskip 5mm \tt
\leftline{DO  xyzt=0,NXYZT2-1} 
\leftline{...}
}
\leftline{The parity of the results is undefined.}
  
If $\tt .Rand.$ has an argument of type real\_field, it returns
a real\_field of pseudorandom numbers uniformely distributed between
0 and the corresponding component of the argument.  The parity of
the result is the same as the parity of the argument.

The unary operator $\tt .Gauss.$ returns a real field of
pseudorandom numbers with gaussian distribution
of mean square deviation equal to the argument of
$\tt .Gauss.$ and upgrades the global variable $\tt seeds$. 
The argument can again be a variable of kind $\tt REAL8$
or of type real\_field and the parity of the result is undefined
or equal to the parity of the argument, respectively. 

The unary operator $\tt .Ggauss.$ works like $\tt .Gauss.$ but fills with
gaussian random numbers the components of a generator\_field, 
setting its direction equal to 0.  Precisely, the instruction
$\tt ge=.Ggauss.r $, although in the module it is implemented 
differently, would be equivalent to

\vskip 4mm
{\baselineskip 5mm \tt
\leftline{DO i=1,8}
\leftline{\quad auxr=.Gauss.r}
\leftline{\quad ge\%fc(i,:,:,:,:)=auxr\%fc}
\leftline{END DO}
\leftline{ge\%parity=auxr\%parity}
\leftline{ge\%dir=0}
}
\leftline{where $\tt auxr$ is a variable of type real\_field}

This module assumes the availability of long (8-byte) integers and the fact
that a multiplication of long integers will return the lowest 8 bytes of the
product (i.e. $a*b\; {\rm mod}\; 2^{64}$) without producing an arithmetic error
when the product exceeds the maximum long integer.  If these assumptions
are not satisfied, the module should be replaced with some other suitable
module.  Also, we would like to point out that the algorithm of 
Eq.~(\ref{erand}) produces pseudorandom numbers of reasonably good quality
and period ($ \approx 10^{14}$).  However, a computer capable of
100 Gflops sustained running a program that makes use of one pseudorandom
number every thousand floating point operations would exhaust the
whole set of pseudorandom numbers in one million seconds, which is not
a very long time.  Thus for calculations that are very computer intensive
or which demand pseudorandom numbers of exceptionally good
quality, the module should be modified to meet the more
stringent demands.  Two improvements which can be made with 
minimal changes would consist in the use of a larger $m$ 
in Eq.~(\ref{erand}) (with appropriate $a_1$ and $b_1$) and/or of
a reshuffle of the pseudorandom numbers produced by the algorithm.
Of course, one could also make use of the F90 RANDOM\_NUMBER
subroutine, but the results would no longer be machine independent.

\subsection{Module conditionals}
\label{conditionals}

This module defines six overloaded relational operators, $\tt >$, $\tt >=$,
$\tt < $, $\tt <= $, $\tt == $, $\tt /=$, and the operator $\tt .Xor.$

The relational operators take as operands two real\_fields or one 
real\_field and one real variable of kind $\tt REAL8$.
They return a logical variable which is set to $\tt .TRUE.$ if 
the two fields have the same (defined) parity or if the single
field operand has defined parity, and is set to $\tt .FALSE.$ otherwise.  
At the same time the global variable $\tt context$ 
is set to $\tt .TRUE.$ at all sites where the relation is satisfied, 
to $\tt .FALSE.$ at all other sites.  For example, the function
implementing the relational operator $\tt a>b$, with
$\tt a$ and $\tt b$ of type real\_field, could contain a line:
$\tt\; context=a\%fc>b\%fc \;$, which produces the action mentioned above.

The operator $\tt .Xor.$ accepts as operands a pair of fields
of the same type and returns a field, also of the same type,
having as elements the corresponding elements of the first operand 
at the sites where the global variable $\tt context$ is $\tt .TRUE.$, 
the elements of the second operand at the sites where $\tt context$
is $\tt .FALSE.$.  For clarification, the function $\tt g\_xor\_g$ 
implementing the operation $\tt g1 .Xor. g2$, where $\tt g1$
and $\tt g2$ are fields of type gauge\_field,  would contain
the code

\vskip 4mm
{\baselineskip 5mm \tt
\leftline{DO j=1,3}
\leftline{DO i=1,3}
\leftline{\quad WHERE(context)} 
\leftline{\quad  \quad g\_xor\_g\%fc(i,j,:,:,:,:)=g1\%fc(i,j,:,:,:,:)}
\leftline{\quad  ELSEWHERE}
\leftline{\quad  \quad g\_xor\_g\%fc(i,j,:,:,:,:)=g2\%fc(i,j,:,:,:,:)}
\leftline{\quad  END WHERE}
\leftline{END DO}
\leftline{END DO}
}        

The $\tt parity$ of the field returned by $\tt .Xor.$ is the common
parity of the two operands if they have the same $\tt parity$,
otherwise it is undefined.  In addition, for operands of type
gauge\_field, the $\tt dir$ component of the returned field is the
common $\tt dir $ of the operands if they have the same $\tt dir$,
otherwise it is set to 0.

The operators provided by the module ``conditionals'' can be very 
conveniently used to select elements out of two fields according
to some local condition, an operation which lies at the foundation
of stochastic simulation techniques. 

\subsection{Subroutine write\_conf and read\_conf}
\label{wrconf}

The file ``write\_read\_conf.f90'' contains two subroutines which
serve to store and retrieve an entire SU(3) gauge field configuration,
written in a portable, compressed ASCII format.  Only the first two 
columns of the gauge field matrices are stored, because
the third one can be recovered from the unitarity and unimodularity
constraints.  The write\_conf subroutine takes advantage of the fact
that all of the elements of the gauge field matrices have magnitude
smaller or equal to 1 to re-express their real and imaginary parts 
as 48bit integers.  These integers are then written in base 64,
with the digits being given by the ASCII collating sequence
starting from 0.  Thus, 8 characters are needed to represent either
the real or the imaginary part of the gauge field matrix elements
and an entire gauge field matrix is represented by 96 ASCII characters,
without loss of numerical information.  A detailed description of
the contents of the file generated by write\_conf and of the 
standard used for coding the information is written, as a header,
at the beginning of the file itself.  This makes the file with
the compressed gauge configuration portable and usable, irrespective
of the availability of the write\_conf and read\_conf subroutines or
of a separate description of their implementation.

\section{Example code}
\label{excode}
In order to illustrate the power of the modules we developed, we
present here two programs which implement a multihit Metropolis
simulation of quenched QCD and the calculation of a quark propagator.
Anybody familiar with the complexity of the programs for
lattice QCD simulations will appreciate the conciseness of our
examples.  It is also to be noticed that a large amount of the
code in the programs performs peripheral functions, such as
initialization and printout.  If we consider the Metropolis
simulation program, for instance, the section of the code which
performs the actual upgrading steps consists of only 28 lines!
It is our hope that researchers interested in using our modules
will find it easy to become familiar with their functionality
and that, not being hindered by inessential programming burdens,
they will thus be able to make much faster progress in the development
of far-reaching QCD applications.


\subsection{quenched.f90}

{\chardef \other = 12
\def\deactivate{%
   \catcode `\\ = \other   \catcode`\{ = \other
   \catcode `\} = \other   \catcode`\$ = \other
   \catcode `\& = \other   \catcode`\# = \other
   \catcode `\% = \other   \catcode`\~ = \other
   \catcode `\^ = \other   \catcode`\_ = \other
}

\def\makeactive#1{\catcode`#1 = \active \ignorespaces}
{%
   \makeactive\^^M %
   \gdef\obeywhitespace{%
      \makeactive\^^M %
      \let^^M = \newline %
      \aftergroup\removebox %
      \obeyspaces %
   }%
}
\def\newline{\par\indent}
\def\removebox{\setbox0=\lastbox}

\def\verbatim{\par\begingroup\deactivate\obeywhitespace\tt\parindent = 0in
\baselineskip=5mm   \catcode `\| = 0 %
}

\def\endverbatim{\endgroup}

\def\|{|}

\verbatim

!  Program Qcdf90_quenched 

!  Copyright by Indranil Dasgupta, Andrea R. Levi, Vittorio Lubicz
!  and Claudio Rebbi  -  Boston University  -  May 1996
!  This program may be freely copied and used as long as this notice
!  is retained.

PROGRAM Qcdf90_quenched

   USE precisions
   USE constants
   USE global_module
   USE field_algebra
   USE generator_algebra
   USE conditionals
   USE shift
   USE random_numbers
   USE assign_mixed
   USE assign_isotype1

   IMPLICIT NONE
   TYPE(gauge_field):: staple,g_old,g_new
   TYPE(real_field):: plaq_old,plaq_new,bf_ratio,rand
   TYPE(generator_field):: ge
   TYPE(matrix) :: zero_matrix, unit_matrix
   LOGICAL l_test,l_seed
   REAL(REAL8) clock_dcl,clock_upd,clock_plaq
   REAL(REAL8) beta,saved_beta,hp,av_plaq,aux,range_small,range_unit
   CHARACTER(LEN=64) in_filename,out_filename
   CHARACTER(LEN=16) id
   INTEGER(LONG) saved_seed,inp_seed
   INTEGER clock_rate,clock_1,clock_2
   INTEGER hotcoldread,save,num_upd,p,m,sign,nu,i,hit,num_hit

! input variables:
   WRITE (*,'("Lattice size: ",4I5)') NX,NY,NZ,NT
   WRITE (*,'("Enter beta:   ")',ADVANCE='NO')
   READ *,beta
   WRITE (*,'("Enter number of updates:   ")',ADVANCE='NO')
   READ *,num_upd
   WRITE (*,'("Select the starting configuration.  Enter 0 for&
               &a hot start ")')
   WRITE (*,'("1 for a cold start, 2 to read from Disk: ")',ADVANCE='NO')
   READ *,hotcoldread

! other useful variables:
   num_hit=6                ! number of Metropolis multiple hits
   range_unit=1._REAL8      ! unitary range for the random numbers
   range_small=0.1_REAL8    ! range for the random numbers
   inp_seed=1               ! input seed for random numbers generator
   zero_matrix
   in_filename= 'configuration.in'    
   out_filename='configuration.out'

! initializing system clock
   CALL SYSTEM_CLOCK(clock_1,clock_rate)
   clock_dcl=1._REAL8/clock_rate
   clock_upd=0._REAL8
   clock_plaq=0._REAL8

! initializing random generator and gauge configuration
   SELECT CASE(hotcoldread)
   CASE(0)
     l_seed=.Seed.inp_seed
     DO p=0,1
     DO m=1,4
       ge=.Ggauss.range_unit
       u
       u
       u
     END DO
     END DO
   CASE(1)
     l_seed=.Seed.inp_seed
     unit_matrix
     DO p=0,1
     DO m=1,4
       u
       u
       u
     END DO
     END DO
   CASE(2)
     CALL read_conf(saved_beta,id,hp,saved_seed,in_filename)    
     IF(inp_seed==0) THEN
       WRITE (*,'("saved_seed=",I15)') saved_seed
       l_seed=.Seed.saved_seed
     ELSE
       l_seed=.Seed.inp_seed
       WRITE (*,'("seed re-initialized")')
     ENDIF
   CASE DEFAULT
     WRITE (*,'("hotcoldread must only be 0,1 or 2")')
     STOP
   END SELECT

   DO i=1,num_upd                        ! Main Loop

! Metropolis update 
      CALL SYSTEM_CLOCK(clock_1)
      DO p=0,1
      DO m=1,4

        ! Staple
        staple=zero_matrix
        staple
        staple
        DO nu=1,4
          IF(nu.EQ.m) CYCLE
          DO sign=-1,1,2
            staple=staple+((nu*sign).Ushift.u
          END DO
        END DO

        g_old=u
        DO hit=1,num_hit
          plaq_old=g_old.Dot.staple
          ge=.Ggauss.range_small
          ge
          ge
          g_new=(.Exp.ge)*g_old
          plaq_new=g_new.Dot.staple
          bf_ratio=.Exp.(beta/3._REAL8*(plaq_new-plaq_old))
          rand=.Rand.range_unit
          l_test=rand<bf_ratio
          assign_type='t'; g_old=g_new
        END DO
        u
      END DO
      END DO
      CALL SYSTEM_CLOCK(clock_2)
      clock_upd=clock_upd+(clock_2-clock_1)*clock_dcl

! Plaquette
      CALL SYSTEM_CLOCK(clock_1)
      av_plaq=0._REAL8
      DO p=0,1
      DO m=1,3
        DO nu=m+1,4
          aux=u
          av_plaq=av_plaq+aux
        END DO
      END DO
      END DO
      av_plaq=av_plaq/REAL(18*NXYZT,REAL8)
      CALL SYSTEM_CLOCK(clock_2)
      clock_plaq=clock_plaq+(clock_2-clock_1)*clock_dcl
     WRITE (*,'("iteration ",I5," av. plaq.= ",F10.6)') i,av_plaq

   END DO                                ! End Main Loop

! Save configuration on disk
   WRITE (*,'("Save configuration on disk ? (Yes=1, &
               &No=0): ")',ADVANCE='NO')
   READ *,save
   IF(save==1) THEN
     WRITE (*,'("saving the configuration")')
     id='conf 0.0.0'
     hp=0.0
     l_seed=.TRUE.
     saved_seed=.Seed.l_seed
     WRITE (*,'("  saved_seed = ",I15)') saved_seed
     CALL write_conf(beta,id,hp,saved_seed,out_filename)
   ENDIF

! Print timing
   WRITE (*,'("Av. upgrade time in microsecs per link",F9.3)') &
            (1000000*clock_upd)/(4*NXYZT*num_upd)
   WRITE (*,'("Av. measure time in microsecs per plaquette",F9.3)')&
            (1000000*clock_plaq)/(6*NXYZT*num_upd)

   END

|endverbatim

\subsection{propagator.f90}
\verbatim

!  Program Qcdf90_propagator

!  Copyright by Indranil Dasgupta, Andrea R. Levi, Vittorio Lubicz
!  and Claudio Rebbi  -  Boston University  -  January 1996
!  This program may be freely copied and used as long as this notice
!  is retained.

PROGRAM Qcdf90_propagator

   USE precisions
   USE constants
   USE global_module
   USE field_algebra
   USE generator_algebra
   USE conditionals
   USE shift
   USE dirac
   USE random_numbers
   USE assign_mixed
   USE assign_isotype1
   USE assign_isotype2

   IMPLICIT NONE
   TYPE(fermi_field):: psi,chi,grad,h,m_h,mp_m_h
   REAL(REAL8) clock_dcl,clock_cg
   REAL(REAL8) kappa,tolerance,residue,saved_beta,hp
   REAL(REAL8) alpha,old_alpha,g_2,g_old_2,beta_cg,old_beta_cg
   REAL(REAL8) h_a_h,norm_psi
   CHARACTER(LEN=64) in_filename
   CHARACTER(LEN=16) id
   INTEGER(LONG) saved_seed
   INTEGER clock_rate,clock_1,clock_2
   INTEGER iter,nsteps,niter,init_niter,stop_flag,init_stop_flag
   INTEGER i,xyzt,s

! input variables:
   WRITE (*,'("Enter kappa:   ")',ADVANCE='NO')
   READ *,kappa
   WRITE (*,'("Enter max numbers of cg steps:   ")',ADVANCE='NO')
   READ *,nsteps
   WRITE (*,'("Enter tolerance:   ")',ADVANCE='NO')
   READ *,tolerance
                              ! the conjugated gradient will
                              ! run until the residue<tolerance
                              ! or for a maximum of nsteps

! other useful variables:
   in_filename= 'configuration.in'
   init_stop_flag=2
   init_niter=4

! gauge configuration is read from the disk
   CALL read_conf(saved_beta,id,hp,saved_seed,in_filename)

! the source chi (in the even sites) is set arbitrarly in this example
   DO i=1,3
   DO s=1,4
   DO xyzt=0,NXYZT2-1
     chi
   END DO
   END DO
   END DO
   chi

! psi must be initialized as the starting trial configuration.
! the simplest choice is psi=chi
   psi=chi

! initializing system clock
   CALL SYSTEM_CLOCK(clock_1,clock_rate)
   clock_dcl=1._REAL8/clock_rate

!  Calculate psi as the solution of: M*psi=chi,
!  where M is the fermion matrix, and chi is a given source.
!  The residue is printed to monitor the convergence.

    stop_flag=init_stop_flag
    niter=init_niter
    iter=0
    m_h=psi-(kappa**2)*(.Dirac.(.Dirac.psi))
    mp_m_h=m_h-(kappa**2)*(.Xdirac.(.Xdirac.m_h))
    grad=chi-mp_m_h
    g_2=grad*grad
    h=grad
    norm_psi=psi*psi
    residue=g_2/norm_psi
    WRITE (*,'("residue= ",F20.16," at step:",I5)') residue,iter
    old_alpha=0._REAL8
    old_beta_cg=1._REAL8

    DO iter=1,nsteps
      m_h=h-(kappa**2)*(.Dirac.(.Dirac.h))
      h_a_h=m_h*m_h
      beta_cg=g_2/h_a_h
      psi=psi+beta_cg*h
      norm_psi=psi*psi
      IF(mod(iter,niter)==0 .AND. g_2/norm_psi<tolerance) THEN
        stop_flag=stop_flag-1
        m_h=psi-(kappa**2)*(.Dirac.(.Dirac.psi))
        mp_m_h=m_h-(kappa**2)*(.Xdirac.(.Xdirac.m_h))
        grad=chi-mp_m_h
        g_2=grad*grad
        h=grad
        g_old_2=g_2
        residue=g_2/norm_psi
        WRITE (*,'("residue= ",F20.16," at step:",I5)') residue,iter
        IF(stop_flag == 0) EXIT
      ELSE
        mp_m_h=m_h-(kappa**2)*(.Xdirac.(.Xdirac.m_h))
        grad=grad-beta_cg*mp_m_h
        g_old_2=g_2
        g_2=grad*grad
        alpha=g_2/g_old_2
        h=grad+alpha*h
        norm_psi=psi*psi
        residue=g_2/norm_psi
        IF(mod(iter,niter) == 0) THEN
           WRITE (*,'("residue= ",F20.16," at step:",I5)') residue,iter
        ENDIF
        old_beta_cg=beta_cg
        old_alpha=alpha
      END IF

    END DO
    CALL SYSTEM_CLOCK(clock_2)
    clock_cg=(clock_2-clock_1)*clock_dcl

!test solution:
   m_h=psi-(kappa**2)*(.Dirac.(.Dirac.psi))
   mp_m_h=m_h-(kappa**2)*(.Xdirac.(.Xdirac.m_h))
   grad=chi-mp_m_h
   norm_psi=psi*psi
   g_2=grad*grad
   residue=g_2/norm_psi
   WRITE (*,'("final residue= ",F20.16)') residue

! Print timing
   WRITE (*,'("Cg time per iteration per link in microsecs",F9.3)') &
         (1000000*clock_cg)/(iter*4*NXYZT)

   END

|endverbatim

\subsection{Compilation and sample run output}

The code has been tested and compiled on a SGI PowerChallengeArray 
with 90 MHz processor nodes, using IRIX 6.1 Fortran 90,
with a single processor -O3 optimization flags or 
with the flags -O3 -pfa -mp to implement multiprocessing;
on a SGI Indigo using the IRIX 6.1 Fortran 90;
and on the IBM R6000 58H model 7013 at 55 MHz, 
with the xlf90 IBM compiler using the -O3 optimization flags.

The run of the example programs produce the following
outputs when running on a single processor of 
the SGI PowerChallengeArray.

{\it Output of quenched.f90}

\verbatim

Lattice size:     8    8    8    8
Enter beta:   6.0
Enter number of updates:   15
Select the starting configuration.  Enter 0 for a hot start 
1 for a cold start, 2 to read from Disk: 1
iteration     1 av. plaq.=   0.849923
iteration     2 av. plaq.=   0.773278
iteration     3 av. plaq.=   0.727001
iteration     4 av. plaq.=   0.699791
iteration     5 av. plaq.=   0.677709
iteration     6 av. plaq.=   0.664358
iteration     7 av. plaq.=   0.654980
iteration     8 av. plaq.=   0.645880
iteration     9 av. plaq.=   0.638568
iteration    10 av. plaq.=   0.635049
iteration    11 av. plaq.=   0.631868
iteration    12 av. plaq.=   0.628131
iteration    13 av. plaq.=   0.624450
iteration    14 av. plaq.=   0.621757
iteration    15 av. plaq.=   0.619540
Save configuration on disk ? (Yes=1, No=0): 1
saving the configuration
  saved_seed = 182618478903297
Av. upgrade time in microsecs per link  275.533
Av. measure time in microsecs per plaquette    8.624

|endverbatim

{\it Output of propagator.f90}

\verbatim

Enter kappa:   0.155
Enter max numbers of cg steps:   2000
Enter tolerance:   1.e-14
residue=   0.2417309784323778 at step:    0
residue=   0.0047398663491857 at step:    4
residue=   0.0007379739612745 at step:    8
residue=   0.0001969957469930 at step:   12
residue=   0.0000640891236947 at step:   16
residue=   0.0000235032623593 at step:   20
residue=   0.0000079006298388 at step:   24
residue=   0.0000030559375109 at step:   28
residue=   0.0000014286464825 at step:   32
residue=   0.0000006891018149 at step:   36
residue=   0.0000003502268064 at step:   40
residue=   0.0000002296348520 at step:   44
residue=   0.0000001023143421 at step:   48
residue=   0.0000000357544848 at step:   52
residue=   0.0000000114945632 at step:   56
residue=   0.0000000034230144 at step:   60
residue=   0.0000000010030646 at step:   64
residue=   0.0000000005153800 at step:   68
residue=   0.0000000003507320 at step:   72
residue=   0.0000000002150572 at step:   76
residue=   0.0000000000880532 at step:   80
residue=   0.0000000000289906 at step:   84
residue=   0.0000000000095210 at step:   88
residue=   0.0000000000042035 at step:   92
residue=   0.0000000000016253 at step:   96
residue=   0.0000000000004389 at step:  100
residue=   0.0000000000001207 at step:  104
residue=   0.0000000000000298 at step:  108
residue=   0.0000000000000066 at step:  112
residue=   0.0000000000000022 at step:  116
final residue=   0.0000000000000022
Cg time per iteration per link in microsecs   21.548

|endverbatim
}

\section*{Acknowledgments}

This research was supported in part under DOE grant DE-FG02-91ER40676.
We are grateful to the Center of Computational Science and the Office
of Information Technology for support and access to the Boston
University supercomputer facility.
V.L. acknowledges the support of an INFN post-doctoral fellowship.

\end{document}